\documentclass[12pt]{article}
\usepackage{latexsym}
\usepackage{amsmath,amsfonts}
\usepackage{times}
\allowdisplaybreaks[4]
\catcode`\@=11

\newcount\hour
\newcount\minute
\newtoks\amorpm \hour=\time\divide\hour by 60\minute
=\time{\multiply\hour by 60 \global\advance\minute by-\hour}
\edef\standardtime{{\ifnum\hour<12 \global\amorpm={am}%
        \else\global\amorpm={pm}\advance\hour by-12 \fi
        \ifnum\hour=0 \hour=12 \fi
        \number\hour:\ifnum\minute<10
        0\fi\number\minute\the\amorpm}}
\edef\militarytime{\number\hour:\ifnum\minute<10 0\fi\number\minute}

\def\draftlabel#1{{\@bsphack\if@filesw {\let\thepage\relax
   \xdef\@gtempa{\write\@auxout{\string
      \newlabel{#1}{{\@currentlabel}{\thepage}}}}}\@gtempa
   \if@nobreak \ifvmode\nobreak\fi\fi\fi\@esphack}
        \gdef\@eqnlabel{#1}}
\def\@eqnlabel{}
\def\@vacuum{}
\def\marginnote#1{}
\def\draftmarginnote#1{\marginpar{\raggedright\scriptsize\tt#1}}
\def\draft{
        \pagestyle{plain}
        \overfullrule=2pt
        \oddsidemargin -.5truein
        \def\@oddhead{\sl \phantom{\today\quad\militarytime} \hfil
        \smash{\Large\sl DRAFT} \hfil \today\quad\militarytime}
        \let\@evenhead\@oddhead
        \let\label=\draftlabel
        \let\marginnote=\draftmarginnote
        \def\ps@empty{\let\@mkboth\@gobbletwo
        \def\@oddfoot{\hfil \smash{\Large\sl DRAFT} \hfil}
        \let\@evenfoot\@oddhead}
        \def\@eqnnum{(\theequation)\rlap{\kern\marginparsep\tt\@eqnlabel}%
        \global\let\@eqnlabel\@vacuum}  }

\newcommand{\rf}[1]{(\ref{#1})}
\renewcommand{\theequation}{\thesection.\arabic{equation}}
\renewcommand{\thefootnote}{\fnsymbol{footnote}}
\newcommand{\newsection}{   % Numeration of eqs. is automatic
\setcounter{equation}{0}\section}

\def\appendix#1{\addtocounter{section}{1}\setcounter{equation}{0}
\renewcommand{\thesection}{\Alph{section}}
\section*{Appendix \thesection\protect\indent \parbox[t]{11.15cm}{#1}}
\addcontentsline{toc}{section}{Appendix \thesection\ \ \ #1}}

\def\be{\begin{equation}}
\def\ee{\end{equation}}
\def\beq{\begin{eqnarray}}
\def\eeq{\end{eqnarray}}

\def\xline{\, x \kern -0.56em /}
\def\parline{\,\partial\kern -0.55em /\,\,}

\def\half{{\frac{1}{2}}}

\def\AA{{\cal A}}
\def\BB{{\cal B}}

\def\LL{{\cal L}}
\def\MM{{\cal M}}

\def\TT{{\cal T}}

\def\Dbf{{\bf D}}
\def\Gbf{{\bf G}}
\def\ebf{{\bf e}}

\def\Dline{{D \kern-0.7em  /  }\,\, }
\def\Dlinebf{{{\bf D} \kern-0.7em  /  }\,\, }
\def\irm{{\rm i}}
\def\cov{{\rm cov}}
\def\d{{\rm d}}
\def\u{{\rm u}}

\def\bos{{\rm bos}}
\def\eff{{\rm eff}}

\def\st{{\rm st}}

\def\mod{{\rm mod}}
\def\on-sh{{\rm on-sh}}
\def\off-sh{{\rm off-sh}}
\def\cf{{\rm cf}}
\def\cur{{\rm cur}}
\def\sh{{\rm sh}}
\def\AdS{{\rm AdS}}

\def\phik{|\phi\rangle}

\def\psik{|\psi\rangle}
\def\psibr{\langle\psi|}
\def\Psik{|\Psi\rangle}
\def\Psibr{\langle\Psi|}

\def\xik{|\xi\rangle}

\def\Xik{|\Xi\rangle}

\def\smzero{{\scriptscriptstyle (0)}}
\def\smone{{\scriptscriptstyle (1)}}

\def\smzero{{\scriptscriptstyle (0)}}
\def\smone{{\scriptscriptstyle (1)}}

\def\smponetwo{{\scriptscriptstyle [1,2]}}
\def\smponethree{{\scriptscriptstyle [1,3]}}

\def\Awt{\widetilde{A}}

\def\Ywt{\widetilde{Y}}

\def\ewt{\widetilde{e}}

\def\Ewh{\widehat{E}}

\def\gaal{\gamma\alpha}
\def\gaalb{\gamma\bar\alpha}
\def\alpar{\alpha\partial}
\def\albpar{\bar\alpha\partial}

\def\Cb{\bar{C}}
\def\eb{\bar{e}}

\def\rb{\bar{r}}

\def\alphabf{{\boldsymbol{\alpha}}}
\def\gammabf{{\boldsymbol{\gamma}}}

\def\gaalbf{\gammabf\alphabf}
\def\gaalbbf{\gammabf\bar\alphabf}

\def\Gammasm{{\scriptscriptstyle{\Gamma}}}

\begin{document}

%\draft

\begin{flushright}
FIAN-TD-2013-18\qquad \ \ \ \ \  \ \\
arXiv: 1311.7350 [hep-th]
\end{flushright}

\vspace{1cm}

\begin{center}

{\Large \bf  CFT adapted approach to massless fermionic fields, AdS/CFT,

\medskip

and fermionic conformal fields}

\vspace{2.5cm}

R.R. Metsaev\footnote{ E-mail: metsaev@lpi.ru }

\vspace{1cm}

{\it Department of Theoretical Physics, P.N. Lebedev Physical
Institute, \\ Leninsky prospect 53,  Moscow 119991, Russia }

\vspace{3.5cm}

{\bf Abstract}

\end{center}

Fermionic totally symmetric
arbitrary spin massless fields in AdS space of dimension greater than or
equal to four are studied. Using Poincar\'e parametrization of AdS space, CFT adapted gauge invariant formulation for such fields is developed. We demonstrate that the curvature and radial coordinate contributions to Lagrangian and gauge transformation of the AdS fields can be expressed in terms of ladder operators. Covariant and modified de Donder gauge conditions are proposed. The modified de Donder gauge leads to decoupled equations of motion which can easily be solved in terms of the Bessel function. The AdS/CFT correspondence for conformal current and shadow field and the respective normalizable and non-normalizable modes of fermionic massless AdS field is studied. The AdS field is considered by using the modified de Donder gauge which simplifies considerably the study of AdS/CFT correspondence. We show that on-shell leftover gauge symmetries of bulk massless field are related to gauge symmetries of boundary conformal current and shadow field. We compute the bulk action on solution of the Dirichlet problem and obtain two-point gauge invariant vertex of shadow field. Also we shown that the UV divergence of the two-point gauge invariant vertex gives higher-derivative action of  fermionic conformal field.

\newpage
\renewcommand{\thefootnote}{\arabic{footnote}}
\setcounter{footnote}{0}

%%%%%%%%%%%%%%%%%%%%%%%%%%%%%%%%%%%%%%%%%%%
\section{Introduction}
%%%%%%%%%%%%%%%%%%%%%%%%%%%%%%

Conjectured duality  of conformal SYM theory
and superstring theory in $AdS_5 \times S^5$ in Ref.\cite{Maldacena:1997re} has lead to intensive and in-depth study of various interrelations between AdS field (string) dynamics and CFT. Interesting approach for the studying interrelation between AdS field (string) theories and the corresponding CFT has been proposed in Refs.\cite{wit,Gubser:1998bc}. In Refs.\cite{wit,Gubser:1998bc} it has been proposed that AdS field (string) theory action evaluated on the solution of field equations of motion with the Dirichlet problem corresponding to the boundary shadow field can be considered as generating function for correlation functions of the corresponding boundary CFT. In this paper,
AdS field theory action evaluated on the solution of
field equations of motion with the Dirichlet problem corresponding
to the boundary shadow field will be referred to as effective action.
Obviously, developing the methods for the computation of effective action for concrete AdS field theories is of crucial importance in studying various aspects of the AdS/CFT correspondence. In general, computation of effective action turns out to be complicated problem. One of ways to simplify analysis of AdS field (string) dynamics, and hence to simplify the computation of effective action, is based on use of the Poincar\'e parametrization of
AdS space%
\footnote{ Studying $AdS_5\times S^5$ superstring action
\cite{Metsaev:1998it} in Poincar\'e parametrization of AdS may be found in
Ref.\cite{Metsaev:2000ds}.}.
As is well known, use of the Poincar\'e coordinates simplifies considerably analysis of many aspects of AdS field dynamics and this is the reason why these coordinates have extensively been used for the computation of effective action. In Refs.\cite{Metsaev:2008ks,Metsaev:2009hp}, we developed a Lagrangian gauge invariant formulation of arbitrary spin bosonic massless and massive AdS fields which  is based on considering of AdS field dynamics in the Poincar\'e coordinates. Because the formulation developed in Refs.\cite{Metsaev:2008ks,Metsaev:2009hp} turns out to be very convenient for the studying AdS/CFT correspondence for arbitrary spin fields (see Refs.\cite{Metsaev:2009ym,Metsaev:2010zu,Metsaev:2011uy}), we refer to this formulation as CFT adapted formulation of AdS field dynamics. The purpose of this paper is to develop CFT adapted formulation for fermionic massless arbitrary spin AdS fields and apply this formulation for the studying AdS/CFT correspondence.
Our results can be summarized as follows.

\noindent {\bf i}) Using the Poincar\'e parametrization of $AdS_{d+1}$ space, we obtain gauge invariant
Lagrangian for fermionic free massless arbitrary spin AdS field.  The Lagrangian is
{\it explicitly invariant with respect to boundary Poincar\'e symmetries},
i.e., manifest symmetries of our Lagrangian are adapted to manifest
symmetries of boundary CFT. We show that all the curvature and radial
coordinate contributions to our Lagrangian and gauge transformation are
entirely expressed in terms of two ladder operators that depend on radial
coordinate and radial derivative. Besides this, our Lagrangian and gauge
transformation are similar to the ones of Stueckelberg formulation of massive
field in flat $d$-dimensional space. General structure of the Lagrangian we
obtained is valid for any theory that respects Poincar\'e symmetries. Various
theories are distinguished by appropriate ladder operators.

\noindent {\bf ii}) We find two gauge conditions which we refer to as modified de Donder gauge and covariant de Donder gauge. Modified de Donder gauge leads to simple gauge-fixed equations of motion. The surprise is that this gauge gives {\it decoupled equations of motion}. To our knowledge, the covariant de Donder gauge for arbitrary spin fermionic fields has not been discussed in the earlier literature. Therefore we present our results for the covariant de Donder gauge for fermionic fields. The covariant de Donder gauge respects Lorentz symmetries of AdS space but leads to coupled equations of motion. In
contrast to this, our modified de Donder gauge leads to simple decoupled
equations which are easily solved in terms of the Bessel function.

\noindent {\bf iii})We use our CFT adapted formulation for the studying AdS/CFT correspondence between fer\-mionic massless arbitrary spin AdS field and the corresponding arbitrary spin boundary
conformal current and shadow field. Namely, we show that non-normalizable modes of arbitrary spin-$(s+\half)$ massless AdS field are related to arbitrary spin-$(s+\half)$ shadow field, while normalizable modes of arbitrary spin-$(s+\half)$ massless AdS field are related to arbitrary spin-$(s+\half)$ conformal current. We recall that, in earlier literature, the AdS/CFT correspondence between non-normalizable modes of spin-$\half$ AdS field and the corresponding spin-$\half$ shadow field was studied in Refs.\cite{Henningson:1998cd,Mueck:1998iz,Arutyunov:1998ve,Henneaux:1998ch}, while the AdS/CFT correspondence between non-normalizable modes of massless spin-$\frac{3}{2}$  AdS field (gravitino field) and the corresponding spin-$\frac{3}{2}$ shadow field was studied in Refs.\cite{Corley:1998qg,Volovich:1998tj,Rashkov:1999ji}.
The AdS/CFT correspondence for spin-$(s+\half)$ massless
AdS field with $s>1$ and the corresponding spin-$(s+\half)$ conformal current and shadow
field has not been considered in the earlier literature.%
\footnote{ Scaling properties of 2-point effective action for arbitrary spin fermionic field in $AdS_5$ were studied in Ref.\cite{Germani:2004jf}.}

\noindent {\bf iv}) To compute 2-point effective action we use the modified de Donder gauge. As we have already said, the modified de Donder gauge leads
to the simple gauge-fixed decoupled bulk equations of motion which are easily solved. These gauge-fixed equations have on-shell leftover bulk gauge symmetries.
We show that these on-shell leftover bulk gauge symmetries are realized as
the gauge symmetries of boundary conformal current and shadow field. This is to say that first-order equations of motion for fermionic fields and modified de Donder gauge lead to differential constraints for conformal current and shadow fields. These differential constraints are invariant under gauge transformation of conformal current and shadow field. We find Lagrangian for massless arbitrary spin fermionic fields with proper boundary term and compute the 2-point effective action. The effective action also turns out to be invariant under gauge transformation of shadow field. We give various representations for the effective action.

\noindent {\bf v}) For the case of bosonic fields, it is well known that UV divergence of effective action coincides with higher-derivative action for conformal bosonic fields  (for spin-2 field, see Ref.\cite{Liu:1998bu},
while for arbitrary spin field, see Ref.\cite{Metsaev:2009ym}).%
\footnote{
For discussion of $N=4$ conformal supergravity, see Refs.\cite{Balasubramanian:2000pq,Buchbinder:2012uh}. In the framework of the AdS/CFT correspondence, recent interesting discussion of conformal fields may be found in Refs.\cite{Giombi:2013yva,Tseytlin:2013jya}.}
We demonstrate that UV divergence of effective action for fermionic fields gives higher-derivative action for conformal fermionic  fields. We obtain two representation for the action of conformal fermionic fields.

\newsection{Gauge invariant action of  fermionic massless AdS field }

In this section, using arbitrary parametrization of AdS space, we review Lagrangian metric-like formulation of fermionic massless arbitrary spin AdS fields developed in Refs.\cite{Fang:1979hq,Metsaev:2006zy}.%
\footnote{ In the framework of metric-like approach, massless fermionic fields in $AdS_4$ were studied in Ref.\cite{Fang:1979hq}, while massless fermionic fields in $AdS_{d+1}$, with $d\geq 3$, were studied in Ref.\cite{Metsaev:2006zy}. Frame-like approach to massless fermionic fields was discussed in Ref.\cite{Vasiliev:1987tk} (see also Ref.\cite{Alkalaev:2001mx}-\cite{Skvortsov:2010nh}). In Ref.\cite{Hallowell:2005np}, massless fermionic fields were studied by using the radial dimensional reduction method in Ref.\cite{Biswas:2002nk}.  In the framework of BRST approach, discussion of fermionic fields may be found in Refs.\cite{Biswas:2002nk}-\cite{Moshin:2007jt}.}
In $(d+1)$-dimensional $AdS_{d+1}$ space, a massless totally symmetric
arbitrary spin fermionic field is labelled by one half-integer spin label $s+\frac{1}{2}$, where $s>0$ is an integer number. To discuss Lorentz covariant and gauge invariant
formulation of such field we introduce Dirac complex-valued
tensor-spinor field of the $so(d,1)$ Lorentz algebra,
\be \label{collect}
\Psi^{A_1\ldots A_s\alpha}\,,
\ee
where $A=0,1,\ldots, d$ are flat vector indices of the $so(d,1)$
algebra.

The tensor-spinor field
$\Psi^{A_1\ldots A_s}$ is subject to the basic algebraic constraint
\be
\label{gamtra1} \gamma^A \Psi^{ABBA_4\ldots A_s} =0 \,,
\ee
which tells us that the tensor-spinor field $\Psi^{A_1\ldots A_s}$ is a
reducible representation of the Lorentz algebra $so(d,1)$%
\footnote{ Constraint \rf{gamtra1} was introduced in Ref.\cite{Fang:1978wz} while study of massless fermionic fields in flat space. This constraint implies that the field
$|\Psi\rangle$ being reducible representation of the Lorentz
algebra $so(d,1)$ is decomposed into spin $s+\frac{1}{2}$,
$s-\frac{1}{2}$, $s-\frac{3}{2}$ irreps of the Lorentz algebra.
Various Lagrangian formulations in terms of unconstrained fields in
flat space and AdS space may be found e.g., in Refs.\cite{Francia:2002aa}-\cite{Francia:2010qp}.}.
Note that for $s=0,1,2$ the constraint \rf{gamtra1} is satisfied
automatically.

In order to obtain the gauge invariant description of the massless field
in an easy--to--use form, let us introduce the creation and
annihilation operators $\alpha^A$ and $\bar{\alpha}^A$
defined by the relations%
\footnote{ We use oscillator formulation to handle the many indices appearing for arbitrary spin fields (see e.g., Refs.\cite{Lopatin:1987hz,Bekaert:2006ix}).}
\be
\label{intver15} [\bar\alpha^A,\,\alpha^B]= \eta^{AB}\,,\qquad
\qquad \bar\alpha^A|0\rangle=0\,,
\ee
where $\eta^{AB}$ is the mostly positive flat metric tensor. The
oscillators $\alpha^A$, $\bar\alpha^A$
transform in the vector representations of the
$so(d,1)$ Lorentz algebra. The tensorial component of the tensor-spinor field \rf{collect}
can be collected into a ket-vector $|\Psi\rangle$ defined by
\be
\label{genfun2} |\Psi\rangle \equiv \frac{1}{s!} \alpha^{A_1}\ldots
\alpha^{A_s} \Psi^{A_1\ldots A_s\alpha}|0\rangle\,.
\ee
Here and below spinor indices of ket-vectors are implicit.  The ket-vector
$\Psik$ \rf{genfun2} satisfies the constraint
\be
\label{homcon2} (N_\alphabf  -s)|\Psi\rangle =0\,,\qquad N_\alphabf \equiv \alpha^A\bar\alpha^A\,,
\ee
which tells us that $|\Psi\rangle$ is a degree-$s$ homogeneous
polynomial in the oscillator $\alpha^A$. We note also, that, in terms of the ket-vector $\Psik$ \rf{genfun2}, the algebraic constraint \rf{gamtra1} takes the form
\beq
&& \label{gamtra1n} \gaalbbf \bar\alphabf^2 |\Psi\rangle =0\,,
\\
&& \gaalbbf \equiv \gamma^A\bar\alpha^A\,,\qquad \bar\alphabf^2 \equiv
\bar\alpha^A\bar\alpha^A\,.
\eeq

Action and Lagrangian for the massless fermionic field in $AdS_{d+1}$ space
take the form
\beq
\label{22112013-man-01} && \hspace{3cm} S = \int d^{d+1} x \, \LL\,,
\\
\label{25112013-man01-01} && \hspace{-1.1cm} {\rm i} e^{-1} \LL =   \Psibr E \Psik\,,
\\
\label{25112013-man01-02}  && \hspace{-1.1cm} E \equiv E_\smone + E_\smzero\,,
\\
&& \hspace{-1.1cm} E_\smone \equiv \Dlinebf  - \alphabf \Dbf \gaalbbf -
\gaalbf \bar\alphabf \Dbf + \gaalbf \Dlinebf \gaalbbf +
\frac{1}{2} \gaalbf \alphabf \Dbf \bar\alphabf^2 +
\frac{1}{2}\alphabf^2 \gaalbbf \bar\alphabf \Dbf -
\frac{1}{4}\alphabf^2\Dlinebf\bar\alphabf^2,
\\
\label{MMmassdef} && \hspace{-1.1cm} E_\smzero \equiv (1 - \gaalbf \gaalbbf - \frac{1}{4}
\alphabf^2 \bar\alphabf^2) \ebf_1^\Gammasm\,,
\\
\label{22112013-man1-04} && \hspace{-1.1cm} \ebf_1^\Gammasm \equiv s + \frac{d-3}{2}\,,
\eeq
where $\Psibr$ is defined according the rule $\Psibr =
(\Psik)^\dagger\gamma^0$. We use $e\equiv \det e_\mu^A$, where
$e_\mu^A$ is vielbein of $AdS_{d+1}$ space. We use the notation
\be
\gaalbf \equiv  \gamma^A\alpha^A\,, \quad \gaalbbf
\equiv \gamma^A\bar\alpha^A\,, \quad \alphabf^2 \equiv \alpha^A
\alpha^A\,,\quad \bar\alphabf^2\equiv\bar\alpha^A \bar\alpha^A \,,
\ee
\be
\Dlinebf \equiv \gamma^A D^A\,,\qquad \alphabf \Dbf \equiv \alpha^A
D^A\,,\qquad \bar\alphabf \Dbf \equiv \bar\alpha^A D^A\,, \qquad
D_A \equiv e_A^\mu D_\mu\,,
\ee
and $e_A^\mu$ stands for inverse vielbein of $AdS_{d+1}$ space, while
$D_\mu$ stands for the Lorentz covariant derivative
\be \label{lorspiope} D_\mu \equiv
\partial_\mu
+\frac{1}{2}\omega_\mu^{AB}M^{AB}\,.\ee
The $\omega_\mu^{AB}$ is the Lorentz connection of $AdS_{d+1}$ space,
while a spin operator $M^{AB}$ forms a representation of the Lorentz
algebra $so(d,1)$:
\beq
\label{loralgspiope} && M^{AB} = M_\bos^{AB} + \frac{1}{2}\gamma^{AB}\,,
\\
&& M_\bos^{AB} \equiv \alpha^A \bar\alpha^B - \alpha^B \bar\alpha^A\,,  \qquad \gamma^{AB} \equiv
\frac{1}{2}(\gamma^A\gamma^B - \gamma^B\gamma^A)\,.
\eeq

Now we discuss gauge symmetries of the action in \rf{22112013-man-01}.
To this end we introduce parameter of gauge transformations
$\Xi^{A_1\ldots A_{s-1}\alpha}$, which is
$\gamma$-traceless (for $s'>0$) Dirac complex-valued tensor-spinor
spin-$(s-\frac{1}{2})$ field of the $so(d,1)$ Lorentz algebra,
\be \label{epscollect}
\Xi^{A_1\ldots A_{s-1}\alpha}\,,\qquad \quad \gamma^A \Xi^{AA_2\ldots
A_{s-1}}=0\,, \quad \hbox{for }\ \  s>0\,. \ee
As before to simplify our expressions we use the ket-vector of gauge
transformations parameter
\beq
\label{gaugpar1}
&& \Xik  \equiv \frac{1}{(s-1)!} \alpha^{A_1}\ldots
\alpha^{A_{s-1}} \Xi^{A_1\ldots A_{s-1}\alpha}|0\rangle\,.
\eeq
The ket-vector $\Xik$ satisfies the algebraic constraints
\beq
&& \label{gaugpar3} (N_\alphabf   - s + 1 )\Xik =0\,,
\\
&& \label{gaugpar4} \gaalbbf \Xik =0\,.
\eeq
The constraint \rf{gaugpar3} tells us that the ket-vector $\Xik$
is a degree-$(s-1)$ homogeneous polynomial in the oscillators
$\alpha^A$, while the constraint \rf{gaugpar4} respects the
$\gamma$-tracelessness of $\Xi^{A_1\ldots A_{s-1}}$ \rf{epscollect}.

Now the gauge transformations under which the action \rf{22112013-man-01} is
invariant take the form
\beq
\label{gaugtrapsi} \delta \Psik & = & \Gbf \Xik \,,
\\
\label{gaugtrapsi-xx} && \Gbf \equiv \alphabf \Dbf + \half \gaalbf\,.
\eeq
%

%%%%%%%%%%%%%%%%%%%%%%%%%%%%%%%%%%%%%%%%%%%%%%%%%%%%%%%%%%%%%%%%%%%%%%%
\subsection{ Covariant de Donder gauge condition }
%%%%%%%%%%%%%%%%%%%%%%%%%%%%%%%%%%%%%%%%%%%%%%%%%%%%%%%%%%%%%%%%%%%%%%%

Below, for the study of AdS/CFT correspondence, we will use modified de Donder gauge. In this section we would like to discuss covariant de Donder gauge condition which to our knowledge has not been discussed in the earlier literature.%
\footnote{ For bosonic fields discussion of the standard de Donder gauge
may be found e.g., in Refs.\cite{Guttenberg:2008qe}. Recent interesting discussion of
modified de Donder gauge may be found in Ref.\cite{Chang:2011mz}. We believe that our
covariant and modified de Donder gauge conditions will also be useful for better understanding of various aspects of AdS/QCD correspondence which are discussed, e.g., in Refs.\cite{Brodsky:2013dca}.}.
To this end we consider equations of motion for the fermionic field $\Psik$ obtained from Lagrangian given in \rf{25112013-man01-01},
\be \label{25112013-man01-03}
E \Psik = 0 \,,
\ee
where operator $E$ is given in  \rf{25112013-man01-02}. It is easy to make sure that equations \rf{25112013-man01-03} amount to the following equations
\beq
\label{25112013-man01-04}  && \Ewh\Psik = 0 \,,
\\
\label{25112013-man01-04x1} && \bar\AA_e \Psik = 0 \,,
\\
\label{25112013-man01-04x2} && \bar\BB_e \Psik = 0 \,,
\\
&& \Ewh \equiv  \Dlinebf - \alphabf \Dbf \gaalbbf + \ebf_1^\Gammasm + \half  \gaalbf\gaalbbf\,,
\\
&& \bar\AA_e \equiv \bar\alphabf \Dbf - \Dlinebf \gaalbbf - \half \alphabf \Dbf \bar\alphabf^2 + (\ebf_1^\Gammasm + \half)\gaalbbf - \frac{1}{4}\gaalbf \bar\alphabf^2\,,
\\
&& \bar\BB_e \equiv \bar\alphabf \Dbf \gaalbbf - \half \Dlinebf \bar\alphabf^2 -\half  (\ebf_1^\Gammasm + 1)\bar\alphabf^2\,,
\eeq
where $\ebf_1^\Gammasm$ is given in \rf{22112013-man1-04}.
Equations \rf{25112013-man01-04} turn out to be more convenient for the derivation of second-order equations. This is to say that by acting with operator $\Dlinebf$ on the l.h.s of equations \rf{25112013-man01-04} we obtain second-order equations for fermionic fields which can be cast into the form
\beq
\label{25112013-man01-05} && \Bigl( \Dlinebf^2 - \Gbf {\bf \Cb}_\cov - \ebf_1^\Gammasm \ebf_1^\Gammasm - \alphabf^2\bar\alphabf^2\Bigr)\Psik =0 \,,
\\
\label{25112013-man01-06} && {\bf \Cb}_\cov \equiv {\bf \Cb}_\st - \half (\gaalbbf + \half \gaalbf \bar\alphabf^2)\,,
\\
\label{25112013-man01-07} && {\bf \Cb}_\st \equiv \bar\alphabf \Dbf - \half \alphabf \Dbf \bar\alphabf^2\,,
\eeq
where operator $\Gbf$ is defined in \rf{gaugtrapsi-xx}. For the reader convenience, we note the relation
\beq
\label{25112013-man01-09} &&  \Dlinebf{}\!^2\Psik  = \Bigl( \Box_\AdS  + s+ \frac{d(d+1)}{4} - \gaalbf \gaalbbf \Bigr) \Psik\,,
\\
&& \Box_{\AdS} \equiv D^A D^A + \omega^{AAB}D^B\,.
\eeq
Gauge invariant equations of motion \rf{25112013-man01-05} motivate us to introduce a gauge condition which we refer to as covariant de Donder gauge
\be \label{25112013-man01-10}
{\bf \Cb}_\cov \Psik = 0 \,,\hspace{3cm} \hbox{covariant de Donder gauge}\,,
\ee
where the operator ${\bf \Cb}_\cov$ is defined in \rf{25112013-man01-06}, \rf{25112013-man01-07}.

Use of the covariant de Donder gauge \rf{25112013-man01-10} in gauge invariant equations \rf{25112013-man01-05} leads to the following second-order gauge-fixed equations
\be \label{25112013-man01-11}
\Bigl( \Dlinebf{}\!^2 - \ebf_1^\Gammasm \ebf_1^\Gammasm - \alphabf^2 \bar\alphabf^2\Bigr) \Psik = 0 \,.
\ee
Thus, we see that the use of covariant de Donder gauge leads to simple gauge-fixed equations of motion. Note however that equations \rf{25112013-man01-11} are coupled equations. Decoupled equations can be obtained by using CFT adapted approach and modified de Donder gauge which we discuss in next Sections.

The following remarks are in order.

\noindent {\bf i)} We note that gauge-fixed second-order equations of motion \rf{25112013-man01-11} have on-shell leftover gauge symmetries. These on-shell leftover gauge symmetries can simply be
obtained from generic gauge symmetries \rf{gaugtrapsi} by the
substitution $\Xik \rightarrow |\Xi_{\rm lf-ov}\rangle$, where the
$|\Xi_{\rm lf-ov}\rangle$ satisfies the following equations of motion:
\be
\Bigl( \Box_\AdS - (s-1)(s+d-1) - \frac{d+1}{4}\Bigr) |\Xi_{\rm lf-ov}\rangle = 0 \,.
\ee

\noindent {\bf ii)} Covariant de Donder gauge condition \rf{25112013-man01-10} respects algebraic constraint \rf{gaugpar4} only on-shell. In other words, the relation $\gaalbbf {\bf \Cb}_\cov\Psik = 0$ is valid only by using first-order equations of motion \rf{25112013-man01-04x2}. This can easily be seen by noticing the relation
\be
\gaalbbf {\bf \Cb}_\cov \Psik = \bar\BB_e\Psik\,.
\ee
It easy to obtain off-shell extension of gauge condition \rf{25112013-man01-10}. This is to say that by considering $\gamma$-traceless part of gauge condition \rf{25112013-man01-10}, we obtain the following off-shell covariant de Donder gauge condition:
\beq
&& {\bf \Cb}_\cov^{\rm off-sh}\Psik =  0\,,
\\
&& {\bf \Cb}_\cov^{\rm off-sh} \equiv {\bf \Cb}_\cov - \frac{1}{2\ebf_1^\Gammasm} \gaalbf \bar\BB_e\,,
\eeq
where ${\bf \Cb}_\cov$ is given in \rf{25112013-man01-06}.

\newsection{ CFT adapted Lagrangian and gauge symmetries} \label{CFTsecAdS/CFTcur}

We now discuss CFT adapted approach to fermionic massless arbitrary spin-$(s+\half)$ AdS field.
To discuss Lorentz covariant and gauge invariant
formulation of such field we introduce Dirac complex-valued
tensor-spinor fields of the $so(d-1,1)$ Lorentz
algebra $\psi^{a_1\ldots a_{s'}\alpha}$, $s'=0,1,\ldots, s$ (where
$a=0,1,\ldots, d-1$ are flat vector indices of the $so(d-1,1)$
algebra), i.e. we start with the following collection of the tensor-spinor fields:%
\footnote{ Fields in \rf{collect-a1} are obtained from the field in \rf{collect} by the invertible transformation.}
\be
\label{collect-a1} \sum_{s'=0}^s \oplus\, \psi^{a_1\ldots
a_{s'}\alpha}\,.
\ee
Note that with respect to the spinor index $\alpha$ we prefer to use nomenclature of $so(d,1)$ algebra. This is to say that the spinor index $\alpha$ takes values $\alpha=1,\ldots, 2^{[(d+1)/2]}$ and the tensor-spinor fields $\psi^{a_1\ldots a_{s'}\alpha}$
can be presented as 2-vectors
\be \label{26112013-man01-01}
\psi^{a_1\ldots a_{s'}\alpha} = \left(
\begin{array}{c}
\psi_\u^{a_1\ldots a_{s'}\alpha}
\\[7pt]
\psi_\d^{a_1\ldots a_{s'}\alpha}
\end{array}
\right)\,,
\ee
where $\psi_\u^{a_1\ldots a_{s'}\alpha}$, $\psi_\d^{a_1\ldots a_{s'}\alpha}$ are subject to constraints
\be
(1-\gamma^z)\psi_\u^{a_1\ldots a_{s'}} = 0 \,, \qquad
(1+\gamma^z)\psi_\d^{a_1\ldots a_{s'}} =0 \,.
\ee
We note also that we use $\gamma$-matrices $\gamma^A = \gamma^a, \gamma^z$ which  are $2^{[(d+1)/2]}\times 2^{[(d+1)/2]}$ matrices of  $so(d,1)$ algebra.

The tensor-spinor fields $\psi^{a_1\ldots a_{s'}\alpha}$ are symmetric with respect to vector indices of the $so(d-1,1)$ algebra $a_1 \ldots a_{s'}$ and subject to the basic algebraic constraints
\be
\label{gamtra1-a1q} \gamma^a \psi^{abba_4\ldots a_{s'}} =0 \,, \qquad s'=3,4,\ldots, s\,.
\ee

In order to obtain the gauge invariant description of a massless field
in an easy--to--use form, let us introduce a set of the creation and
annihilation operators $\alpha^a$, $\alpha^z$ and $\bar{\alpha}^a$,
$\bar\alpha^z$ defined by the relations%
\be\label{intver15-a1}
[\bar\alpha^a,\,\alpha^b]= \eta^{ab}\,,\qquad
[\bar\alpha^z,\,\alpha^z] = 1\,,
\qquad \bar\alpha^a|0\rangle=0\,,\qquad \bar\alpha^z|0\rangle=0\,,\ee
where $\eta^{ab}$ is the mostly positive flat metric tensor. The
oscillators $\alpha^a$, $\bar\alpha^a$ and $\alpha^z$, $\bar\alpha^z$
transform in the respective vector and scalar representations of the
$so(d-1,1)$ Lorentz algebra. Tensorial components of the tensor-spinor fields \rf{collect-a1} can be collected into a ket-vector $|\psi\rangle$ defined by
\beq
\label{intver16n1-a1}
&& |\psi\rangle  \equiv \sum_{s'=0}^s \frac{\alpha_z^{s-s'}}{\sqrt{(s-s')!}}
|\psi^{s'}\rangle\,,
\\
&& \label{genfun2-a1} |\psi^{s'}\rangle \equiv \frac{1}{s'!} \alpha^{a_1}\ldots
\alpha^{a_{s'}} \psi^{a_1\ldots a_{s'}\alpha}|0\rangle\,.
\eeq
The ket-vectors $|\psi^{s'}\rangle$ \rf{genfun2-a1} satisfy the constraints
\beq
\label{homcon2-a1} && (N_\alpha -s')|\psi^{s'} \rangle =0\,,
\qquad s'=0,1,\ldots, s\,,\qquad  N_\alpha \equiv \alpha^a\bar\alpha^a\,,
\\
\label{gamtra1-a1} && \gamma\bar\alpha \bar\alpha^2 |\psi^{s'} \rangle
=0\,, \qquad \qquad s' = 3,4,\ldots, s\,.
\eeq
Constraints \rf{homcon2-a1} tell us that $|\psi^{s'}\rangle$ is a degree-$s'$ homogeneous
polynomial in the oscillator $\alpha^a$, while  constraints \rf{gamtra1-a1} amount to the ones in \rf{gamtra1-a1q}. Note that for $s'=0,1,2$ the constraints \rf{gamtra1-a1} are satisfied automatically. In terms of the ket-vector $\psik$ \rf{intver16n1-a1}, the
algebraic constraints \rf{homcon2-a1},\rf{gamtra1-a1} take the form
\beq
&& \label{homcon2n-a1} (N_\alpha  + N_z  - s )|\psi\rangle =0\,,\qquad  N_\alpha \equiv \alpha^a\bar\alpha^a\,, \qquad N_z \equiv \alpha^z\bar\alpha^z\,,
\\
&& \label{gamtra1n-a1} \gamma\bar\alpha \bar\alpha^2 |\psi\rangle =0\,.
\eeq
Equation \rf{homcon2n-a1} tells us that $|\psi\rangle$ is a degree-$s$
homogeneous polynomial in the oscillators $\alpha^a$, $\alpha^z$.

Using Poincar'e parametrization of AdS, we find CFT adapted action and Lagrangian for the massless fermionic field in $AdS_{d+1}$,
\beq
\label{action-a1} && \hspace{3cm} S = \int d^d x dz\, \LL\,,
\\
\label{22112013-man1-05} && \hspace{-1cm} {\rm i} \LL =   \psibr E \psik\,,
\\
\label{28112013-man1-04} && E \equiv E_\smone + E_\smzero\,,
\\
\label{22112013-man1-05-c1} && E_\smone \equiv
\parline  - \alpha\partial\gamma\bar\alpha -
\gamma\alpha\bar\alpha\partial + \gamma\alpha
\parline\gamma\bar\alpha + \frac{1}{2}\gamma\alpha\alpha\partial\bar\alpha^2
+ \frac{1}{2}\alpha^2\gamma\bar\alpha\bar\alpha\partial -
\frac{1}{4}\alpha^2\parline\bar\alpha^2\,, \ \ \ \
\\
&& E_\smzero = (1 - \gamma\alpha\gamma\bar\alpha -
\frac{1}{4}\alpha^2\bar\alpha^2) e_1^\Gammasm + (\gamma\alpha -
\half\alpha^2\gamma\bar\alpha)\eb_1
+ (\gamma\bar\alpha -\half \gamma\alpha\bar\alpha^2) e_1\,,
\eeq

\beq
\label{21112013-man1-01} && e_1^\Gammasm =  e_{1,1}^\Gammasm \bigl( \sigma_- \TT_{\nu-\half} - \TT_{-\nu+\half} \sigma_+\bigr)\,,
\\
\label{21112013-man1-02} && e_1 = e_{1,1} \TT_{\nu-\half}\,, \qquad \eb_1 =  \TT_{-\nu + \half} \eb_{1,1}\,,
\\
\label{22112013-man1-03} && e_{1,1} = -\alpha^z \ewt_1 \,, \qquad \eb_{1,1} =  - \ewt_1 \bar\alpha^z\,,\qquad \ewt_1 = \Bigl(\frac{2s+d-3-N_z}{2s+d-4-2N_z}\Bigr)^{1/2}\,,
\\
\label{21112013-man1-03-y1} && e_{1,1}^\Gammasm = \frac{2s+d-2}{2s+d-2-2N_z}\,,
\\
&& \TT_\nu \equiv \partial_z + \frac{\nu}{z}\,,\qquad \parline \equiv \gamma^a\partial^a\,, %
\\
\label{21112013-man1-03-x1}  && \nu \equiv s + \frac{d-3}{2} - N_z + \half \sigma_3\,,
\eeq
where $\sigma_3$ is the Pauli matrix, while $2\times 2$ matrices $\sigma_\pm$ are defined in  \rf{manold-03112011-01}. We note that operator $e_1^\Gammasm$ \rf{21112013-man1-01} can be represented as
\be
e_1^\Gammasm =  \sigma_- e_{1,1}^\Gammasm \Bigl(\partial_z + \frac{1}{z}(s+ \frac{d-3}{2}-N_z)\Bigr) + \sigma_+ e_{1,1}^\Gammasm \Bigl(-\partial_z + \frac{1}{z}(s+ \frac{d-3}{2}-N_z)\Bigr)\,.
\ee

{\bf Gauge symmetries}. Now we discuss gauge symmetries of the action given in \rf{action-a1}. To this end we introduce parameters of gauge transformations
$\xi^{a_1\ldots a_{s'}\alpha}$, $s'=0,1,\ldots, s-1$ which are
$\gamma$-traceless (for $s'>0$) Dirac complex-valued tensor-spinor
fields of the $so(d-1,1)$ Lorentz algebra, i.e.,
we start with a collection of the tensor-spinor fields
\be
\label{epscollect-a1} \sum_{s'=0}^{s-1} \oplus\, \xi^{a_1\ldots
a_{s'}\alpha}\,,\qquad \quad \gamma^a \xi^{aa_2\ldots
a_{s'}}=0\,, \quad \hbox{for }\ \  s'>0\,.
\ee
As in \rf{26112013-man01-01}, we assume that the parameter $\xi^{a_1\ldots a_{s'}\alpha}$ is presented as 2-vector
\be \label{26112013-man01-02}
\xi^{a_1\ldots a_{s'}\alpha} = \left(
\begin{array}{c}
\xi_\u^{a_1\ldots a_{s'}\alpha}
\\[7pt]
\xi_\d^{a_1\ldots a_{s'}\alpha}
\end{array}
\right)\,,
\ee
\be
(1-\gamma^z)\xi_\u^{a_1\ldots a_{s'}} = 0 \,, \qquad
(1+\gamma^z)\xi_\d^{a_1\ldots a_{s'}} =0 \,.
\ee
As before to simplify our expressions we use the ket-vector of gauge
transformations parameter
\beq
\label{gaugpar1-a1}
&& \xik  \equiv \sum_{s'=0}^{s-1}
\frac{\alpha_z^{s-1-s'}}{\sqrt{(s-1-s')!}} |\xi^{s'}\rangle\,,
\\
&& \label{gaugpar2-a1} |\xi^{s'}\rangle \equiv \frac{1}{s'!} \alpha^{a_1}\ldots
\alpha^{a_{s'}} \xi^{a_1\ldots a_{s'}\alpha}|0\rangle\,.
\eeq
The ket-vector $\xik$ satisfies the algebraic constraints
\beq
&& \label{gaugpar3-a1} (N_\alpha  + N_\zeta  - s + 1 )\xik =0\,,
\\
&& \label{gaugpar4-a1} \gamma\bar\alpha \xik =0\,.
\eeq
The constraint \rf{gaugpar3-a1} tells us that the ket-vector $\xik$
is a degree-$(s-1)$ homogeneous polynomial in the oscillators
$\alpha^a$, $\alpha^z$, while the constraint \rf{gaugpar4-a1} respects the
$\gamma$-tracelessness of $\xik$.

Now the gauge transformations under which the action \rf{action-a1} is
invariant take the form

\beq
\label{22112013-man1-06} && \delta \psik = G\xik\,,
\\
\label{22112013-man1-07} && \hspace{1cm} G \equiv \alpha\partial - e_1 + \gaal
\frac{1}{2N_\alpha+d-2}e_1^\Gammasm - \alpha^2
\frac{1}{2N_\alpha+d}\eb_1 \,,
\eeq
where operators $e_1$, $\eb_1$, $e_1^\Gammasm$ are given in \rf{21112013-man1-01}, \rf{21112013-man1-02}.

%%%%%%%%%%%%%%%%%%%%%%%%%%%%%%%%%%%%%%%%%%%%%%%%%%%%%%%%%%%%%%%%%%%%%%%%%%%%%%%%%
%%%%%%%%%%%%%%%%%%%%%%%%%%%%%%%%%%%%%%%%%%%%%%%%%%%%%%%%%%%%%%%%%%%%%%%%%%%%%%%%%
\subsection{Global $so(d,2)$ symmetries of CFT adapted action }
%%%%%%%%%%%%%%%%%%%%%%%%%%%%%%%%%%%%%%%%%%%%%%%%%%%%%%%%%%%%%%%%%%%%%%%%%%%%%%%%%
%%%%%%%%%%%%%%%%%%%%%%%%%%%%%%%%%%%%%%%%%%%%%%%%%%%%%%%%%%%%%%%%%%%%%%%%%%%%%%%%%

Relativistic symmetries of $AdS_{d+1}$ space are described by the $so(d,2)$
algebra. In CFT adapted approach, the fermionic massless spin-$(s+\half)$ AdS field is described by the set of the $so(d-1,1)$ algebra fields given in \rf{collect-a1}.
Therefore it is reasonable to represent the $so(d,2)$ algebra so that to
respect manifest $so(d-1,1)$ symmetries. For application to the AdS/CFT
correspondence, most convenient form of the $so(d,2)$ algebra that respects
the manifest $so(d-1,1)$ symmetries is provided by nomenclature of the
conformal algebra. This is to say that the $so(d,2)$ algebra consists of
translation generators $P^a$, conformal boost generators $K^a$, dilatation
generator $D$, and generators $J^{ab}$ which span $so(d-1,1)$ algebra. We use
the following normalization for commutators of the $so(d,2)$ algebra
generators:
\beq
\label{ppkk}
&& {}[D,P^a]=-P^a\,, \hspace{2cm}  {}[P^a,J^{bc}]=\eta^{ab}P^c -\eta^{ac}P^b
\,,
\\
&& [D,K^a]=K^a\,, \hspace{2.2cm} [K^a,J^{bc}]=\eta^{ab}K^c - \eta^{ac}K^b\,,
\\[5pt]
\label{pkjj} && \hspace{2.5cm} {}[P^a,K^b]=\eta^{ab}D - J^{ab}\,,
\\
&& \hspace{2.5cm} [J^{ab},J^{ce}]=\eta^{bc}J^{ae}+3\hbox{ terms} \,.
\eeq
Requiring $so(d,2)$ symmetries implies that the action is invariant with
respect to transformation $\delta_{\hat{G}} \phik  = \hat{G} \phik$, where
the realization of $so(d,2)$ algebra generators $\hat{G}$ in terms of
differential operators takes the form%
\footnote{ In our approach only $so(d-1,1)$ symmetries are realized
manifestly. The $so(d,2)$ symmetries could be realized manifestly by using
ambient space approach (see e.g. Refs.\cite{Metsaev:1995re}-\cite{Bekaert:2009fg}).}
\beq
\label{conalggenlis01} && P^a = \partial^a \,,
\qquad
J^{ab} = x^a\partial^b -  x^b\partial^a + M^{ab}\,,
\\[3pt]
\label{conalggenlis03} && D = x\partial  + \Delta\,,
\qquad
\Delta \equiv  z\partial_z + \frac{d}{2}\,,
\\[3pt]
\label{conalggenlis04} && K^a = -\frac{1}{2}x^2\partial^a + x^a D + M^{ab}x^b
+ R^a \,,
\eeq
$x\partial\equiv x^a\partial^a$, $x^2\equiv x^ax^a$. In
\rf{conalggenlis01},\rf{conalggenlis04}, $M^{ab}$ is spin operator of the
$so(d-1,1)$ algebra. Commutation relations for $M^{ab}$ and representation of
$M^{ab}$ on space of ket-vector $\psik$ \rf{intver16n1-a1} take the form
\beq
&& [M^{ab},M^{ce}]=\eta^{bc}M^{ae}+3\hbox{ terms} \,,
\\
&& M^{ab} = M_\bos^{ab} + \half \gamma^{ab}\,,
\\
&& M_\bos^{ab} \equiv \alpha^a \bar\alpha^b - \alpha^b \bar\alpha^a\,, \qquad \gamma^{ab} \equiv \half (\gamma^a\gamma^b - \gamma^b\gamma^a)\,.
\eeq
Operator $R^a$ appearing in $K^a$ \rf{conalggenlis04} is given by
\beq
&& R^a =  R_\smzero^a  + R_\smone^a\,,
\\
&& R_\smzero^a = r_{0,1}^\Gammasm \Ywt^a + r_{0,1}\bar\alpha^a + \rb_{0,1}\Awt^a\,,
\\
&& R_\smone^a = r_{1,1} \partial^a\,,
\eeq

\beq
&& \Awt^a \equiv \alpha^a - \gaal \gamma^a \frac{1}{2N_\alpha+d-2} -
\alpha^2 \frac{1}{2N_\alpha+d} \bar\alpha^a\,,
\\
&& \Ywt^a \equiv \gamma^a -\gaal \frac{2}{2N_\alpha + d-2}\bar\alpha^a\,,
\\
&& r_{0,1}^\Gammasm = - \frac{\rm i}{2} z e_{1,1}^\Gammasm\sigma_2\,,
\qquad
r_{0,1} = z e_{1,1}\,, \qquad  \rb_{0,1} = - z \eb_{1,1}\,,\qquad r_{1,1} = -\half z^2 \,,
\eeq
where $e_{1,1}$, $\eb_{1,1}$, $e_{1,1}^\Gammasm$ are given in \rf{22112013-man1-03}, \rf{21112013-man1-03-y1}. We note the following interesting relations between operators  $e_1$, $\eb_1$, $e_1^\Gammasm$ and $r_{0,1}$, $\rb_{0,1}$, $r_{0,1}^\Gammasm $,
\be \label{24072013-10}
r_{0,1}^\Gammasm = \half [r_{1,1},e_1^\Gammasm]\,, \qquad \rb_{0,1}
= [\eb_1, r_{1,1}]\,,\qquad r_{0,1} = [r_{1,1}, e_1]\,.
\ee
We see that realization of Poincar\'e symmetries on bulk AdS fields \rf{conalggenlis01}
coincide with realization of Poincar\'e symmetries on boundary CFT operators.
Note that realization of $D$- and $K^a$-symmetries on bulk AdS fields
\rf{conalggenlis03},\rf{conalggenlis04} coincides, by module of contributions
of operators $\Delta$ and $R^a$, with the realization of $D$- and
$K^a$-symmetries on boundary CFT operators. Realizations of the $so(d,2)$
algebra on bulk AdS fields and boundary CFT operators are distinguished by
$\Delta$ and $R^a$. The realization of the $so(d,2)$ symmetries on bulk AdS
fields given in \rf{conalggenlis01}-\rf{conalggenlis04} turns out to be very
convenient for studying the AdS/CFT correspondence.

%%%%%%%%%%%%%%%%%%%%%%%%%%%%%%%%%%%%%%%%%%%%%%%%%%%%%%%%%%%%%%%%%%%%%%%
%%%%%%%%%%%%%%%%%%%%%%%%%%%%%%%%%%%%%%%%%%%%%%%%%%%%%%%%%%%%%%%%%%%%%%%
\subsection{  Modified de Donder gauge}\label{ModgaugesecAdS/CFT}
%%%%%%%%%%%%%%%%%%%%%%%%%%%%%%%%%%%%%%%%%%%%%%%%%%%%%%%%%%%%%%%%%%%%%%%
%%%%%%%%%%%%%%%%%%%%%%%%%%%%%%%%%%%%%%%%%%%%%%%%%%%%%%%%%%%%%%%%%%%%%%%

We begin with discussion of the various forms of equations of motion obtained from Lagrangian given in \rf{22112013-man1-05}. First of all we note that, because operator $E$ \rf{28112013-man1-04} respects the constraint given in \rf{gamtra1n-a1}, equations of motion obtained from \rf{22112013-man1-05} take the form
\be
\label{12072013-04} E \psik= 0 \,.
\ee
It is easy to check that equations \rf{12072013-04} amount to the following equations
\beq
\label{12072013-09}  &&  \widehat{E}\psik  = 0 \,,
\\
\label{12072013-05} && \bar\AA_e\psik = 0 \,,
\\
\label{12072013-06} && \bar\BB_e\psik = 0 \,,
\eeq
where we use the notation
\beq
\label{12072013-01} \widehat{E} & = & \parline - G \gaalb + \Bigl( 1 + \gaal \frac{2}{2N_\alpha+d-2}\gaalb\Bigr)e_1^\Gammasm -\gamma\alpha \frac{2}{2N_\alpha+d-2}\eb_1\,,
\\
\label{12072013-01-x1} \bar\AA_e & = & \bar\AA   + \frac{1}{2N_\alpha+d-2} \Bigl( (2N_\alpha+d-1)\gaalb - \half \gaal \bar\alpha^2 \Bigr)e_1^\Gammasm
\nonumber\\
& + &  \half \bar\alpha^2 e_1 - \frac{1}{2N_\alpha+d-2} \Bigl(2N_\alpha+d - 2 \gaal \gaalb - \half \alpha^2 \bar\alpha^2\Bigr) \eb_1\,,
\\
\label{12072013-01-x2} \bar\BB_e & = &  \bar\BB - \frac{2N_\alpha+d+2}{2(2N_\alpha+d)} \bar\alpha^2 e_1^\Gammasm
+ \frac{1}{2N_\alpha+d} \Bigl(\gaal \bar\alpha^2 - (2N_\alpha+d-2)\gaalb\Bigr) \eb_1\,,
\\
&& \bar\AA \equiv \albpar -  \parline \gaalb - \half
\alpar\bar\alpha^2 \,,
\\
&& \bar\BB \equiv \albpar \gaalb - \half \parline \bar\alpha^2\,.
\eeq

We now discuss second-order equations for fermionic fields. To
this end we note that gauge invariant first-order equations \rf{12072013-04}
lead to the following gauge invariant second-order equations of motion:
\beq
\label{12072013-02}  && \Bigl( \Box  - \MM^2 - G \Cb_\mod^\on-sh\Bigr)\psik = 0 \,,
\\
\label{12072013-02x} && \MM^2 \equiv -\partial_z^2 + \frac{1}{z^2}(\nu^2 - \frac{1}{4})\,,
\\
\label{23112013-man1-14} && \Cb_\mod^\on-sh \equiv \Cb_\st - \frac{1}{2N_\alpha+d-2}(\gaalb + \half \gaal \bar\alpha^2) e_1^\Gammasm + \half \bar\alpha^2 e_1 - \frac{2N_\alpha+d-4}{2N_\alpha+d-2}\Pi_\bos^\smponetwo \eb_1,\qquad
\\
\label{23112013-man1-14-q1}&& \Cb_\st \equiv \albpar - \half \alpar\bar\alpha^2\,,
\eeq
where the operators $\nu$ and $G$ are given in \rf{21112013-man1-03-x1} and \rf{22112013-man1-07} respectively, while a projector $\Pi_\bos^\smponetwo$ is defined in \rf{26112013-man01-04}. Second-order equations \rf{12072013-02} motivate us to introduce a gauge condition which we refer to as {\it modified de Donder gauge condition},
\be \label{22112013-man1-08}
\Cb_\mod^\on-sh \psik = 0 \,, \hspace{3cm} \hbox{modified de Donder gauge},
\ee
where operator $\Cb_\mod^\on-sh $ is given in \rf{23112013-man1-14}, \rf{23112013-man1-14-q1}. Using the modified de Donder gauge
condition \rf{22112013-man1-08} in gauge invariant equations of motion \rf{12072013-02} leads to the following surprisingly simple gauge-fixed equations of motion:
\beq
\label{22112013-man1-09} && \Box_\nu\psik=0 \,,
\\
\label{22112013-man1-09-x1} && \Box_\nu \equiv \Box + \partial_z^2 - \frac{1}{z^2}(\nu^2 - \frac{1}{4})\,,
\eeq
where $\nu$ is defined in \rf{21112013-man1-03-x1}. In terms of tensor-spinor fields
\rf{collect-a1}, equations \rf{22112013-man1-09} can be represented as
\beq
\label{22112013-man1-10} && \Bigl( \Box + \partial_z^2 - \frac{1}{z^2}(\nu_{s'}^2
- \frac{1}{4})\Bigr)\psi^{a_1\ldots a_{s'}} = 0\,,
\\
&& \nu_{s'} \equiv  s' + \frac{d-3}{2} + \half \sigma_3\,,
\eeq
$s'=0,1,\ldots, s$. Thus, our {\it modified de Donder gauge condition
\rf{22112013-man1-08} leads to decoupled equations of motion} \rf{22112013-man1-10} which
can easily be solved in terms of the Bessel function.%
\footnote{ Appearance of Bessel function in the solution of equations of motion for fields in arbitrary background is discussed in Ref.\cite{Gover:2011rz}.
}

The following remarks are in order.

\noindent {\bf i)} We note that gauge-fixed second-order equations of motion \rf{22112013-man1-09} have on-shell leftover gauge symmetries. These on-shell leftover gauge symmetries can simply be
obtained from generic gauge symmetries \rf{22112013-man1-06} by the
substitution $\xik \rightarrow |\xi_{\rm lf-ov}\rangle$, where the ket-vector
$|\xi_{\rm lf-ov}\rangle$ satisfies the following equations of motion:
\be \label{23112013-man1-16}
\Box_\nu|\xi_{\rm lf-ov}\rangle = 0 \,,
\ee
where $\Box_\nu$ is given in \rf{22112013-man1-09-x1}.

\noindent {\bf ii)} We note that our modified de Donder gauge condition \rf{22112013-man1-08} respects the Poincar\'e and dilatation symmetries but breaks the conformal boost $K^a$-symmetries, i.e., the simple form of gauge-fixed equations of motion \rf{22112013-man1-09} is achieved at the cost of the $K^a$-symmetries.

\noindent {\bf iii)} Modified de Donder gauge condition \rf{22112013-man1-08} respects algebraic constraint \rf{gaugpar4-a1} only on-shell. In other words, relation $\gaalb \Cb_\mod^\on-sh\psik = 0$ is valid only by using first-order equations of motion \rf{12072013-06}. This can easily be seen by noticing the relation
\be
\gaalb \Cb_\mod^\on-sh \psik = \bar\BB_e\psik\,.
\ee
It easy to obtain off-shell extension of gauge condition \rf{22112013-man1-08}. This is to say that by considering $\gamma$-traceless part of gauge condition \rf{22112013-man1-08}, we obtain the following off-shell modified de Donder gauge condition
\beq
\label{22112013-man1-11} && \Cb_\mod^\off-sh \psik  = 0 \,, \hspace{3cm} \hbox{off-shell modified de Donder gauge},
\\
\label{28112013-man1-06} && \Cb_\mod^\off-sh \equiv \Cb_\mod^\on-sh - \gaal \frac{1}{2N_\alpha+d} \bar\BB_e\,.
\eeq
Note that operator $\Cb_\mod^\off-sh$ \rf{28112013-man1-06} can be represented as
\beq
&& \Cb_\mod^\off-sh  = (1-\gaal \frac{1}{2N_\alpha+d}\gaalb)\albpar - \half (\alpar - \gaal \frac{1}{2N_\alpha+d}\parline)\bar\alpha^2
\nonumber\\
&& \hspace{1cm} - \frac{1}{2N_\alpha+d-2}(\gaalb
- \gaal \frac{1}{2N_\alpha+d} \bar\alpha^2) e_1^\Gammasm
+ \half \bar\alpha^2 e_1 - \frac{2N_\alpha+d-4}{2N_\alpha+d-2}\Pi^\smponethree \eb_1,\qquad
\eeq
where a projector $\Pi^\smponethree$ is defined in \rf{26112013-man01-03}.

\newsection{AdS/CFT correspondence. Preliminaries}\label{secAdS/CFT}

We now study the AdS/CFT correspondence for free fermionic arbitrary spin massless AdS field and boundary arbitrary spin conformal current and shadow field. To study the AdS/CFT
correspondence we use the gauge invariant CFT adapted formulation of massless AdS field and
modified de Donder gauge condition we discussed in Sections \ref{CFTsecAdS/CFTcur}. We emphasize that it is the use of our massless gauge fields and the modified de Donder gauge condition that leads to the surprisingly simple decoupled gauge-fixed equations of motion%
\footnote{ Discussion of interesting
methods for solving AdS field equations of motion without gauge fixing may be found in
Refs.\cite{Bolotin:1999fa}.}. The use of our massless gauge fields and the modified de Donder
gauge condition makes the study of AdS/CFT correspondence for fermionic arbitrary spin-$(s+\half)$ massless AdS field similar to the one for fermionic  spin-$\half$ massive AdS field. Owing these properties of our massless gauge fields and the modified de Donder gauge condition, the computation of effective action is considerably simplified. Perhaps, this is the main advantage of our approach.

In our approach to the AdS/CFT correspondence, we have gauge symmetries not
only at AdS side but also at the boundary CFT.%
\footnote{ For the first time, the gauge invariant approach to bosonic conformal current and shadow fields was developed in Ref.\cite{Metsaev:2008fs}, while for anomalous conformal current and shadow field such approach was obtained in Refs.\cite{Metsaev:2010zu,Metsaev:2011uy}. In these references,  by using the AdS/CFT correspondence, we also demonstrated how the gauge invariant approach is related to AdS field dynamics. Note however that, before Refs.\cite{Metsaev:2010zu,Metsaev:2011uy} gauge invariant approach to anomalous conformal current was obtained in Refs.\cite{Gover:2008sw,Gover:2008pt} in the framework of tractor approach. Recent interesting discussion of tractor approach may be found in Refs.\cite{Grigoriev:2011gp}.}
Also, we note that the modified de Donder gauge condition turns out to be invariant under on-shell leftover gauge symmetries of massless AdS field. This is to say that, in the
framework of our approach, the study of AdS/CFT correspondence implies the
matching of:
\\
{\bf i}) the bulk first-order equations of motion, the modified de Donder gauge condition for bulk massless field and the corresponding differential constraints for boundary conformal current and shadow field;%
\footnote{ In the framework of standard CFT, discussion of
differential constraints for conformal currents may be
found e.g., in Refs.\cite{Shaynkman:2004vu}-\cite{Alkalaev:2012rg}.}
\\
{\bf ii}) on-shell leftover gauge symmetries of bulk massless field and
the corresponding gauge symmetries of boundary conformal current and shadow field;
\\
{\bf iii}) AdS field action evaluated on the solution of AdS massless
field equations of motion with the Dirichlet problem corresponding
to the boundary shadow field and the boundary two-point
gauge invariant vertex for the shadow field.

\subsection{ AdS/CFT correspondence for massive spin-$\half$ field}

As we have already said, the use of our massless gauge fields and the modified de Donder gauge makes the study of AdS/CFT correspondence for arbitrary spin-$(s+\half)$ massless AdS field similar to the one for spin-$\half$ massive AdS field. Therefore, for the reader convenience, we now briefly recall the AdS/CFT correspondence for the spin-$\half$ massive AdS field.

{\bf AdS/CFT correspondence for normalizable
modes of spin-$\half$ massive AdS field and spin-$\half$ conformal current}.
The action of massive spin-$\half$ field in $AdS_{d+1}$ background takes the form
\beq
\label{19072009-01} && S =  \int d^dx dz\,  \LL \,,
\\
\label{19072009-02} && \irm \LL =  e \bar\Psi (\Dlinebf + m) \Psi\,, \qquad \Dlinebf \equiv \gamma^A D^A\,, \qquad m > 0\,,
\eeq
where $\Psi=\Psi^\alpha$, $\alpha=1,\ldots, 2^{[(d+1)/2]}$,  is Dirac complex-valued spinor field of $so(d,1)$ algebra.

In the CFT adapted approach, we use complex-valued spinor field $\psi= \psi^\alpha$, where as above the spinor index $\alpha$ takes the values $\alpha=1,\ldots, 2^{[(d+1)/2]}$ and the field $\psi$ can be presented as 2-vector
\be \label{26112013-man01-26}
\psi = \left(
\begin{array}{c}
\psi_\u
\\[7pt]
\psi_\d
\end{array}
\right)\,, \qquad (1-\gamma^z)\psi_\u = 0 \,, \qquad
(1+\gamma^z)\psi_\d =0 \,.
\ee

In terms of the field $\psi$, Lagrangian \rf{19072009-02} takes the form (up to total derivative)
\beq
\label{15072013-05} \irm \LL & = & \bar\psi (\parline + e_1^\Gammasm)\psi\,,
\\
\label{23112013-man1-08} &&   e_1^\Gammasm = \sigma_- \TT_{\nu -\half} -
\TT_{-\nu+\half} \sigma_+\,,
\\
\label{23112013-man1-05} && \nu = m + \half \sigma_3\,,
\\
\label{09072009-04} && \TT_\nu \equiv \partial_z +
\frac{\nu}{z}\,.
\eeq
We note that operator $e_1^\Gammasm$ \rf{23112013-man1-08} can be represented as
\be
e_1^\Gammasm = \sigma_- (\partial_z + \frac{m}{z})
+ \sigma_+ (-\partial_z + \frac{m}{z})\,.
\ee
The equation of motion obtained from Lagrangian \rf{15072013-05} is
given by
\be \label{19072009-06}
(\parline + e_1^\Gammasm)\psi  = 0 \,.
\ee
It is easy to check that first-order equation \rf{19072009-06} leads to the following second-order equation
\beq
\label{23112013-man1-07} && \Box_\nu\psi=0 \,,
\\
&& \Box_\nu \equiv \Box + \partial_z^2 - \frac{1}{z^2}(\nu^2 - \frac{1}{4})\,.
\eeq
The normalizable solution of Eq.\rf{23112013-man1-07} takes the form
\beq
\label{man01-16112010-01} && \psi(x,z) = U_\nu^{\rm sc} \psi_\cur(x)\,,
\\
\label{man01-01112010-31x} && \hspace{1cm} U_\nu^{\rm sc} \equiv h_{\nu_0 -1} \sqrt{zq}
J_\nu(zq) q^{-(\nu + \half)}\,,
\\
&& \hspace{1cm} h_\nu\equiv 2^\nu\Gamma(\nu+1)\,,\qquad  q^2\equiv
\Box\,,\qquad
\\
\label{23112013-man1-06} && \hspace{1cm} \nu_0 = m + \half\,,
\eeq
where $J_\nu$ stands for the Bessel function. The asymptotic behavior of
solution \rf{man01-16112010-01} is given by
\be \label{man01-16112010-02}
\psi(x,z) \ \ \stackrel{z\rightarrow 0}{\longrightarrow} \ \ \frac{2^{\nu_0-1}\Gamma(\nu_0)}{2^\nu\Gamma(\nu+1)} z^{\nu + \half}
\psi_\cur(x)\,.
\ee
From \rf{man01-16112010-02}, we see that the field $\psi_\cur$ is indeed the asymptotic
boundary value of the normalizable solution.

Note that the representation for AdS field as 2-vector in \rf{26112013-man01-26} implies the corresponding representation for spin-$\half$ conformal current,
\be \label{26112013-man01-25}
\psi_\cur = \left(
\begin{array}{c}
\psi_{\cur,\u}
\\[5pt]
\psi_{\cur,\d}
\end{array}\right)\,.
\ee
Using representation of dilatation symmetry on space of AdS field in \rf{conalggenlis03} and solution in \rf{man01-16112010-01}, we find realization of the operator of conformal dimension on space of $\psi_\cur$,
\be \label{26112013-man01-27}
\Delta_\cur  = \frac{d+1}{2} + \nu\,, \qquad \nu = m+\half \sigma_3\,.
\ee
From \rf{26112013-man01-27}, we see that conformal dimensions of the fields $\psi_{\cur,\u}$ and $\psi_{\cur,\d}$ are given by
\be
\Delta_\cur(\psi_{\cur,\u}) = \frac{d+2}{2} + m\,, \qquad
\Delta_\cur(\psi_{\cur,\d}) = \frac{d}{2} + m\,.
\ee
Note that the choice of normalization factor $h_{\nu_0 -1}$ in  \rf{man01-01112010-31x} is
a matter of convenience. Our normalization condition implies the following
normalization of asymptotic behavior of the solution in \rf{man01-16112010-01},
\be
\psi_\d(x,z) \ \ \stackrel{z\rightarrow 0}{\longrightarrow} \ \  z^{\nu_0 - \half}
\psi_{\cur,\d}(x)\,.
\ee

{\bf Matching of bulk equation of motion and boundary constraint}. We
now demonstrate how differential constraint for the spin-$\half$ conformal
current is obtained from bulk equation of motion.  To this end we note the relations for Bessel functions, $J_\nu = J_\nu(z)$,
\be \label{man01-01112010-38}
\TT_\nu J_{\nu } = J_{\nu-1}\,,
\qquad
\TT_{-\nu} J_{\nu } = - J_{\nu + 1}\,.
\ee
Using \rf{man01-01112010-38} and solution in \rf{man01-16112010-01} we find the following relation
\be \label{23112013-man1-09}
(\parline + e_1^\Gammasm) \psi(x,z)  =  U_\nu^{\rm sc} (\parline + \sigma_+ \Box + \sigma_-) \psi_\cur(x)\,.
\ee
From \rf{23112013-man1-09}, we see that equation of motion \rf{19072009-06} amounts to the following differential constraint for spin-$\half$ conformal current
\beq
\label{26112013-man01-24} && (\parline + e_{1,\cur}^\Gammasm) \psi_\cur(x) = 0\,,
\\
&& e_{1,\cur}^\Gammasm \equiv \sigma_+ \Box + \sigma_- \,.
\eeq
Using representation of $\psi_\cur$ as 2-vector in \rf{26112013-man01-25},
we see that constraint \rf{26112013-man01-24} allows us to express $\psi_{\cur,\u}$ in terms of $\psi_{\cur,\d}$,
\be
\psi_{\cur,\u} = -\parline \psi_{\cur,\d}\,.
\ee
In other  words, solution to \rf{26112013-man01-24} is given by%
\footnote{ Constraint \rf{26112013-man01-24} and solution \rf{26112013-man01-24-b1} were obtained in the framework of tractor approach in Ref.\cite{Shaukat:2009hp}. Our study demonstrates how the constraint \rf{26112013-man01-24} gives rise in the framework of AdS/CFT correspondence.}
\be \label{26112013-man01-24-b1}
\psi_\cur = \left(
\begin{array}{c}
-\parline \psi_{\cur,\d}
\\[5pt]
\psi_{\cur,\d}
\end{array}\right)\,.
\ee

{\bf AdS/CFT correspondence for non-normalizable modes of spin-$\half$ massive AdS field and spin-$\half$ shadow field}.%
\footnote{ See also Refs.\cite{Henningson:1998cd}-\cite{Henneaux:1998ch}.}
The non-normalizable solution of
Eq.\rf{19072009-06} with the Dirichlet problem corresponding to the boundary shadow field
$\psi_\sh(x)$ can be presented as

\beq
\label{22112013-man1-014} \psi(x,z)  &  =  &  n_\nu \int d^dy\, G_\nu(x-y,z) \psi_\sh(y)\,,
\\
&& n_\nu \equiv   \frac{\Gamma(\nu)}{2^{\pi_-}\Gamma(\nu_0)} \,,
\\
\label{22112013-man1-015} && G_\nu(x,z) = \frac{c_\nu z^{\nu+\half}}{ (z^2+ |x|^2)^{\nu + \frac{d}{2}}
}\,,
\\
\label{22112013-man1-015-b5} && c_\nu \equiv \frac{\Gamma(\nu+\frac{d}{2})}{ \pi^{d/2} \Gamma(\nu)} \,.
\eeq
The asymptotic behaviors of Green function
\rf{22112013-man1-015} and solution \rf{22112013-man1-014} are given by,
\beq
\label{22112013-man1-016} && G_\nu(x,z) \ \ \ \stackrel{z \rightarrow 0}{\longrightarrow} \ \ \ z^{-\nu
+ \half} \delta^d(x)\,,
\\[5pt]
\label{22112013-man1-017} && \psi(x,z) \,\,\, \stackrel{z\rightarrow 0 }{\longrightarrow}\,\,\, z^{-\nu
+ \half} n_\nu \psi_\sh(x)\,.
\eeq
Relation \rf{22112013-man1-017} tells us that solution \rf{22112013-man1-014} has indeed asymptotic behavior corresponding to the shadow field.

Note that representation of AdS fields as 2-vector in \rf{26112013-man01-26} implies the corresponding representation of spin-$\half$ shadow field,
\be \label{26112013-man01-25x1}
\psi_\sh = \left(
\begin{array}{c}
\psi_{\sh,\u}
\\[5pt]
\psi_{\sh,\d}
\end{array}\right)\,.
\ee
Using representation of dilatation symmetry on space of AdS field in \rf{conalggenlis03} and solution in \rf{22112013-man1-014}, we find realization of the operator of conformal dimension on space of $\psi_\sh$,
\be \label{26112013-man01-28}
\Delta_\sh  = \frac{d+1}{2} - \nu\,, \qquad \nu = m+\half \sigma_3\,.
\ee
From \rf{26112013-man01-28}, we see that conformal dimensions of the fields $\psi_{\sh,\u}$ and $\psi_{\sh,\d}$ are given by
\be
\Delta_\sh(\psi_{\sh,\u}) = \frac{d}{2} - m\,, \qquad
\Delta_\sh(\psi_{\sh,\d}) = \frac{d+2}{2} - m\,.
\ee
Note that the choice of normalization factor $n_\nu$ in  \rf{22112013-man1-014} is
a matter of convenience. Our normalization condition implies the following
normalization of asymptotic behavior of the solution in \rf{22112013-man1-014},
\be
\psi_\u(x,z) \ \ \stackrel{z\rightarrow 0}{\longrightarrow} \ \  z^{-\nu_0 + \half}
\psi_{\sh,\u}(x)\,.
\ee

{\bf Matching of bulk equation of motion and boundary constraint}. We
now demonstrate how differential constraint for the spin-$\half$ shadow
field is obtained from bulk equation of motion.  To this end we note the following relations %
\be
\label{23112013-man1-10} (\parline + e_1^\Gammasm) \psi(x,z)   =   n_\nu \int d^dy\, G_\nu(x-y,z) (\parline + \sigma_+ + \sigma_-\Box) \psi_\sh(y)\,.
\ee
From \rf{23112013-man1-10}, we see that equation of motion \rf{19072009-06} amounts to the following differential constraint for spin-$\half$ shadow field
\beq
\label{26112013-man01-29} && (\parline + e_{1,\sh}^\Gammasm) \psi_\sh(x) = 0\,,
\\
&& e_{1,\sh}^\Gammasm \equiv  \sigma_+ + \sigma_-\Box\,.
\eeq
Using representation of $\psi_\sh$ as 2-vector in \rf{26112013-man01-25x1},
we see that constraint \rf{26112013-man01-29} allows us to express $\psi_{\sh,\d}$ in terms of $\psi_{\sh,\u}$,
\be
\psi_{\sh,\d} = -\parline \psi_{\sh,\u}\,.
\ee
In other  words, solution to \rf{26112013-man01-29} is given by
\be \label{26112013-man01-32}
\psi_\sh = \left(
\begin{array}{c}
\psi_{\sh,\u}
\\[5pt]
-\parline \psi_{\sh,\u}
\end{array}\right)\,.
\ee

{\bf Matching of effective action and boundary two-point vertex}. To find the
effective action we follow the standard procedure. Namely, we plug
non-normalizable solution of the bulk equation of motion with the Dirichlet
problem corresponding to the boundary shadow field
\rf{22112013-man1-014} into the bulk action for AdS massless field. In other words we are going to compute effective action defined by the relation
\be \label{23112013-man1-03}
\Gamma_\eff \equiv \int d^dxdz\,  \LL_{\rm on-shell}\,,
\ee
where $ \LL_{\rm on-shell}$ is the Lagrangian evaluated on non-normalizable solution of the bulk equation of motion with the Dirichlet problem corresponding to the boundary shadow field \rf{22112013-man1-014}. Note however that Lagrangian presented in \rf{15072013-05} does not involve a proper boundary term. Expression for Lagrangian involving proper boundary term is given by%
\footnote{ Expression for $\LL$ in \rf{23112013-man1-01} is obtained by straightforward application of methods in Refs.\cite{Arutyunov:1998ve,Henneaux:1998ch}.}
\beq
\label{23112013-man1-01} && \irm \LL = \bar\psi (\gamma^a \overset{\leftrightarrow}\partial{}^a + \overset{\leftrightarrow}e{}_1^\Gammasm)\psi\,,
\\
\label{23112013-man1-02} &&   \overset{\leftrightarrow}e{}_1^\Gammasm = \sigma_- \overset{\rightarrow}\TT_{\nu -\half} +
\overset{\leftarrow}\TT{}_{\nu-\half} \sigma_+\,,
\\
&& \nu = m + \half \sigma_3\,,
\\
\hspace{1cm} && \overset{\leftrightarrow}\partial{}^a \equiv \half (\overset{\rightarrow}\partial{}^a -  \overset{\leftarrow}\partial{}^a)\,, \qquad \overset{\rightarrow}\TT{}_\nu \equiv \overset{\rightarrow}\partial{}_z +
\frac{\nu}{z}\,,  \qquad \overset{\leftarrow}\TT{}_\nu \equiv \overset{\leftarrow}\partial{}_z +
\frac{\nu}{z}\,.
\eeq
We note that operator $\overset{\leftrightarrow}e{}_1^\Gammasm$ \rf{23112013-man1-02} can be represented as
\be
\overset{\leftrightarrow}e{}_1^\Gammasm = \sigma_- (  \overset{\rightarrow}\partial{}_z + \frac{m}{z}) + \sigma_+ ( \overset{\leftarrow}\partial{}_z + \frac{m}{z})\,.
\ee
It is easy to see that Lagrangian \rf{23112013-man1-01} differs from the one in \rf{15072013-05} by total derivatives $\partial^a$, $\partial_z$.

Lagrangian \rf{23112013-man1-01} evaluated on the solution of first-order equations of motion \rf{19072009-06} takes the form
\be
\irm \LL_{\rm on-shell} = \half \partial_z(\bar\psi\sigma_1\psi)\,,
\ee
where $\sigma_1$ is the Pauli matrix. This implies that effective action defined as in \rf{23112013-man1-03} takes the form%
\footnote{ As usually, since solution of the Dirichlet problem
\rf{22112013-man1-014} tends to zero as $z\rightarrow \infty$, we ignore
contribution to $\Gamma_\eff$ \rf{23112013-man1-04} when $z=\infty$. Note that throughout this paper we use conventions corresponding to the Lorentz signature. For the computation of the effective action, we should use the Euclidean signature. All that is required to cast our results into the form corresponding to the Euclidean signature is to make the following replacement for the effective action: $\irm \Gamma_\eff^{\rm Lorentz} \rightarrow \Gamma_\eff^{\rm Euclid}$.}
\be \label{23112013-man1-04}
-\irm \Gamma_\eff \equiv \half \int d^dx\, \psi(x,z) \sigma_1 \psi(x,z) \Bigr|_{z\rightarrow 0}\,.
\ee
Now, plugging in \rf{23112013-man1-04} solution to the second-order equations of motion \rf{22112013-man1-014}, we obtain the following three equivalent representations for the effective action:

\noindent {\bf 1st representation for the effective action}
\beq
\label{26112013-man01-30} -\irm \Gamma_\eff  & = & \nu_0 c_{\nu_0} \int d^d x_1d^d x_2\,  \psi_\sh(x_1) \Bigl( \sigma_+ \frac{f_\nu}{|x_{12}|^{2\nu+d}} + \frac{f_\nu}{|x_{12}|^{2\nu+d}} \sigma_- \Bigr) \psi_\sh(x_2) \,,
\\
&& f_\nu \equiv \frac{\Gamma(\nu+1)\Gamma(\nu+\frac{d}{2})}{ 4^{\nu_0-\nu} \Gamma(\nu_0+1)\Gamma(\nu_0+\frac{d}{2})}\,,
\eeq

\noindent {\bf 2nd representation for the effective action}
\beq
\label{26112013-man01-31} -\irm \Gamma_\eff  & = & \frac{c_{\nu_0}}{4(\nu_0+\frac{d}{2}-1)} \int d^d x_1d^d x_2 \psi_\sh(x_1) \frac{g_\nu}{|x_{12}|^{2\nu+d-2}}(\sigma_+   + \sigma_-\Box) \psi_\sh(x_2) \,,
\\
&& g_\nu \equiv \frac{\Gamma(\nu)\Gamma(\nu+\frac{d}{2}-1)}{ 4^{\nu_0-\nu} \Gamma(\nu_0)\Gamma(\nu_0+\frac{d}{2}-1)}\,,
\eeq
where $\nu$, $\nu_0$, and $c_\nu$ are given in \rf{23112013-man1-05}, \rf{23112013-man1-06}, and \rf{22112013-man1-015-b5} respectively, while the field $\psi_\sh$ is subject to differential constraint in \rf{26112013-man01-29}.

Note also that using constraint \rf{26112013-man01-29}, we can represent \rf{26112013-man01-31} in terms of the Dirac operator,

\noindent {\bf 3rd representation for the effective action}
\be
\label{28112013-man01-02} \irm \Gamma_\eff   =  \frac{c_{\nu_0}}{4(\nu_0+\frac{d}{2}-1)} \int d^d x_1d^d x_2 \psi_\sh(x_1) \frac{g_\nu}{|x_{12}|^{2\nu+d-2}}\parline \psi_\sh(x_2) \,.
\ee

Representations for the effective action given in \rf{26112013-man01-30}, \rf{26112013-man01-31}, \rf{28112013-man01-02} is our solution to problem of 2-point effective action for the case of spin-$\half$ AdS field. Advantage of these representation is that, these representations have straightforward generalization to the case of arbitrary spin fields.  For the case of spin-$\half$ field we consider here, these representations can straightforwardly be related to the one discussed in the earlier literature. All that is required is to plug solution to differential constraint \rf{26112013-man01-32} into our representations. Doing so, we get result in Refs.\cite{Henningson:1998cd}-\cite{Henneaux:1998ch},
\be \label{29082013-01}
-\irm \Gamma_\eff = c_{\nu_0} \int d^d x_1 d^dx_2\, \bar\psi_{\sh,\u}(x_1) \frac{\xline_{12}}{|x_{12}|^{2\nu_0 + d}} \psi_{\sh,\u}(x_2)\,.
\ee

\newsection{ AdS/CFT correspondence for normalizable
modes of massless AdS field and conformal current}\label{secAdS/CFTcur}

We now ready to consider the AdS/CFT correspondence for the spin-$(s+\half)$ massless AdS field and
spin-$(s+\half)$ conformal current. We begin with the discussion of the normalizable
solution of Eq.\rf{22112013-man1-09}. The normalizable solution of Eq.\rf{22112013-man1-09}
is given by
\beq \label{man01-15102011-28}
&& |\psi(x,z)\rangle = U_\nu |\psi_\cur(x)\rangle \,,
\\
\label{man01-15102011-29} && \hspace{1cm} U_\nu \equiv h_{\nu_s -1} (-)^{N_z}
\sqrt{zq} J_\nu(zq) q^{-(\nu + \half)}\,,
\\
\label{man01-26012012-16} && \hspace{1cm} h_\kappa\equiv
2^\kappa\Gamma(\kappa+1)\,,\qquad  q^2\equiv \Box\,, \qquad
\eeq
where we do not show explicitly the dependence of $U_\nu$ on $z$, $q$, and
$\kappa$. The asymptotic
behavior of solution \rf{man01-15102011-28} takes the form
\beq
\label{man01-15102011-30} && |\psi(x,z)\rangle \ \ \stackrel{z\rightarrow 0}{\longrightarrow} \ \ z^{\nu + \half} \frac{2^{\nu_s-1} \Gamma(\nu_s)}{2^\nu\Gamma(\nu+1)}(-)^{N_z}|\psi_\cur(x)\rangle\,,
\\
\label{26112013-man01-05}  && \hspace{3cm} \nu_s \equiv s + \frac{d-2}{2}\,.
\eeq
From \rf{man01-15102011-30}, we see that $|\psi_\cur\rangle$ is indeed boundary value of the normalizable solution.

Note that representation of AdS fields as 2-vectors in \rf{26112013-man01-01}  implies the corresponding representation of fields of spin-$(s+\half)$ conformal current,
\be \label{26112013-man01-34}
\psi_\cur^{a_1\ldots a_{s'}\alpha} = \left(
\begin{array}{c}
\psi_{\cur,\u}^{a_1\ldots a_{s'}\alpha}
\\[7pt]
\psi_{\cur,\d}^{a_1\ldots a_{s'}\alpha}
\end{array}
\right)\,, \qquad s'=0,1,\ldots,s\,.
\ee
Using representation of dilatation symmetry on space of AdS field in \rf{conalggenlis03} and solution in \rf{man01-15102011-28}, we find realization of the operator of conformal dimension on space of $|\psi_\cur\rangle$,
\be \label{26112013-man01-35}
\\
\Delta_\cur = s + d-1 - N_z + \half \sigma_3\,.
\ee
From \rf{26112013-man01-35}, we see that conformal dimensions of the fields in \rf{26112013-man01-34} are given by
\be \label{26112013-man01-41}
\Delta_\cur(\psi_{\cur,\u}^{a_1\ldots a_{s'}\alpha}) = s' + d - \half\,, \qquad
\Delta_\cur(\psi_{\cur,\d}^{a_1\ldots a_{s'}\alpha}) = s' + d - \frac{3}{2}\,.
\ee
Note that the choice of normalization factor $h_{\nu_s -1}$ in  \rf{man01-15102011-29} is
a matter of convenience. Our normalization condition implies the following
normalization of asymptotic behavior of the solution for leading rank-$s$ tensor-spinor field in \rf{man01-16112010-01},
\be
\psi_\d^{a_1\ldots a_s}(x,z) \ \ \stackrel{z\rightarrow 0}{\longrightarrow} \ \  z^{\nu_s - \half}
\psi_{\cur,\d}^{a_1\ldots a_s }(x)\,,
\ee
where $\nu_s$ is given in \rf{26112013-man01-05}.

Now we are going to prove the following statements:

\noindent {\bf i}) For normalizable solution \rf{man01-15102011-28}, the first-order equations of motion \rf{12072013-09} and modified
de Donder gauge condition \rf{22112013-man1-08} lead to the differential
constraints of the spin-$(s+\half)$ conformal current.

\noindent {\bf ii}) On-shell leftover gauge transformation
\rf{22112013-man1-06} of normalizable solution \rf{man01-15102011-28} leads
to the gauge transformation of the spin-$(s+\half)$
conformal current%
\footnote{ Note that gauge transformation given in \rf{22112013-man1-06}
is off-shell gauge transformation. On-shell leftover gauge transformation
is obtained from  gauge transformation \rf{22112013-man1-06} by using gauge transformation
parameter which satisfies equation \rf{23112013-man1-16}.}.

To prove these statements we use the following relations for
the operator $U_\nu$:
\beq
\label{man01-01112010-35} && \TT_{\nu-\half} U_\nu  = U_{\nu-1}\,,
\\[5pt]
\label{man01-01112010-36} && \TT_{-\nu-\half} U_\nu  = - U_{\nu+1}\Box\,,
\eeq
which, in turn, can be derived by using the textbook identities
for the Bessel function given in \rf{man01-01112010-38}.

{\bf Matching of bulk modified de Donder gauge and boundary constraint}. We
now demonstrate how differential constraints for the conformal current are obtained from first-order equations of motion \rf{12072013-09} and modified de Donder gauge condition \rf{22112013-man1-08}. Using \rf{man01-01112010-35} and
\rf{man01-01112010-36}, we find the important relations
\be \label{man01-15102011-60}
e_1^\Gammasm U_\nu = U_\nu e_{1,\cur}^\Gammasm\,,  \qquad e_1 U_\nu = U_\nu e_{1,\cur}\,,  \qquad \eb_1
U_\nu = U_\nu \eb_{1,\cur}\,,
\ee
where operators $e_{1,\cur}^\Gammasm$, $e_{1,\cur}$, $\eb_{1,\cur}$ are given below in \rf{24112013-man1-02}-\rf{24112013-man1-04}.
Acting with operators $\Ewh$ \rf{12072013-01} and $\Cb_\mod^\on-sh$ \rf{23112013-man1-14} on solution $|\psi\rangle$ \rf{man01-15102011-28} and using \rf{man01-15102011-60}, we obtain the relations
\beq
\label{man01-15102011-34x} && \Ewh |\psi(x,z)\rangle = U_\nu \Ewh_\cur |\psi_\cur(x)\rangle\,,
\\
\label{man01-15102011-34} && \Cb_\mod^\on-sh |\psi(x,z)\rangle = U_\nu \Cb_\cur |\psi_\cur(x)\rangle\,,
\eeq
where operators $\Ewh_\cur$ and $\Cb_\cur$ take the form
\beq
\label{24112013-man1-01} && \hspace{-1cm}  \Ewh_\cur \equiv \parline - G_\cur \gaalb + \Bigl( 1 + \gaal \frac{2}{2N_\alpha+d-2}\gaalb\Bigr)e_{1,\cur}^\Gammasm -\gamma\alpha \frac{2}{2N_\alpha+d-2}\eb_{1,\cur}\,,
\\
\label{23112013-man1-15} && \hspace{-1cm} \Cb_\cur \equiv \Cb_\st - \frac{1}{2N_\alpha+d-2}(\gaalb + \half \gaal \bar\alpha^2) e_{1,\cur}^\Gammasm + \half \bar\alpha^2 e_{1,\cur} - \frac{2N_\alpha+d-4}{2N_\alpha+d-2}\Pi_\bos^\smponetwo \eb_{1,\cur},\qquad
\\
\label{23112013-man1-15-n1} && G_\cur \equiv \alpha\partial - e_{1,\cur} + \gaal
\frac{1}{2N_\alpha+d-2}e_{1,\cur}^\Gammasm - \alpha^2 \frac{1}{2N_\alpha+d}\eb_{1,\cur} \,,
\\
&& \Cb_\st \equiv \albpar - \half \alpar\bar\alpha^2\,,
\\
\label{24112013-man1-02} && e_{1,\cur}^\Gammasm =  e_{1,1}^\Gammasm \bigl( \sigma_- + \sigma_+ \Box \bigr)\,,
\\
\label{24112013-man1-03} && e_{1,\cur} = \alpha^z \ewt_1\,,  \qquad \eb_{1,\cur} =  -  \Box \ewt_1 \bar\alpha^z\,,
\\
\label{24112013-man1-04} && e_{1,1}^\Gammasm = \frac{2s+d-2}{2s+d-2-2N_z}\,, \qquad \ewt_1 = \Bigl(\frac{2s+d-3-N_z}{2s+d-4-2N_z}\Bigr)^{1/2}\,.
\eeq
From \rf{man01-15102011-34x}, \rf{man01-15102011-34}, we see that first-order equations of motion \rf{12072013-09} and our modified de Donder gauge condition \rf{22112013-man1-08} lead indeed to the differential constraints for the conformal current given by%
\footnote{ For the case of spin-$\frac{3}{2}$ conformal current,  constraints \rf{26112013-man01-38}, \rf{26112013-man01-39} were obtained in the framework of tractor approach in Ref.\cite{Shaukat:2009hp}. Our discussion provides generalization of the constraints to the case of arbitrary spin conformal current and demonstrates how constraints \rf{26112013-man01-38}, \rf{26112013-man01-39} give rise in the framework of AdS/CFT correspondence.}
\beq
\label{26112013-man01-38} && \Ewh_\cur |\psi_\cur\rangle = 0\,,
\\
\label{26112013-man01-39} && \Cb_\cur |\psi_\cur\rangle = 0\,.
\eeq
As a side of remark we note that constraint \rf{26112013-man01-38} can be represented as
\beq
\label{27112013-man01-01} && E_\cur |\psi_\cur\rangle = 0\,,
\\
\label{27112013-man01-02} && E_\cur \equiv E_{\cur, \smone} + E_{\cur,\smzero}
\\
\label{27112013-man01-03} &&  E_{\cur,\smone} \equiv  E_\smone\,,
\\
\label{27112013-man01-04} && E_{\cur,\smzero} \equiv (1 - \gamma\alpha\gamma\bar\alpha -
\frac{1}{4}\alpha^2\bar\alpha^2) e_{1,\cur}^\Gammasm + (\gamma\alpha -
\half\alpha^2\gamma\bar\alpha) \eb_{1,\cur}
+ (\gamma\bar\alpha -\half \gamma\alpha\bar\alpha^2)  e_{1,\cur}\,,\qquad
\eeq
where the Fang-Fronsdal operator $E_\smone$ appearing in \rf{27112013-man01-03} is given in \rf{22112013-man1-05-c1}.

{\bf Matching of bulk and boundary gauge symmetries}. We now show how leftover
gauge transformation of the massless AdS field is related to gauge transformation of the conformal current. To this end we note
that on-shell leftover gauge transformation of massless AdS field is obtained from
\rf{22112013-man1-06} by plugging gauge transformation parameter, which satisfies equation
\rf{23112013-man1-16}, into \rf{22112013-man1-06}. The normalizable solution of equation for
the gauge transformation parameter \rf{23112013-man1-16} takes the form
\be
\label{man01-15102011-35}  |\xi(x,z)\rangle = U_\nu |\xi_\cur(x)\rangle\,,
\ee
where $U_\nu$ is given in \rf{man01-15102011-29}. On the one hand, plugging
\rf{man01-15102011-35} into \rf{22112013-man1-06} and using
\rf{man01-15102011-60}, we find that bulk on-shell leftover gauge
transformation takes the form
\be \label{man01-15102011-62}
\delta |\psi(x,z)\rangle = U_\nu G_\cur|\xi_\cur(x)\rangle
\,,
\ee
where $G_\cur$ is given in \rf{23112013-man1-15-n1}. On the other hand, relation \rf{man01-15102011-28} leads to
\be \label{man01-15102011-63}
\delta|\psi(x,z)\rangle = U_\nu \delta|\psi_\cur(x)\rangle \,.
\ee
Comparing \rf{man01-15102011-62} and \rf{man01-15102011-63}, we
see that gauge transformation of the conformal current takes the form
\be \label{26112013-man01-40}
\delta |\psi_\cur\rangle =  G_\cur|\xi_\cur\rangle\,.
\ee
We check that differential constraints \rf{26112013-man01-38}, \rf{26112013-man01-39} are invariant under gauge transformation \rf{26112013-man01-40}. Thus we see that the on-shell
leftover gauge symmetries of solution of the Dirichlet problem for the spin-$(s+\half)$ massless AdS field are indeed lead to gauge symmetries of the spin-$(s+\half)$ conformal current.

The following remark is in order.

From gauge transformation \rf{26112013-man01-40}, we learn that some fields in \rf{26112013-man01-34} transform as Stueckelberg fields. These Stueckelberg fields can be gauged away. After this, using differential constraints \rf{26112013-man01-38}, \rf{26112013-man01-39}, we can express all the remaining fields in \rf{26112013-man01-34} in terms of one rank-$s$ tensor-spinor field $\psi_{\cur,\d}^{a_1\ldots a_s}$. Besides this, the field $\psi_{\cur,\d}^{a_1\ldots a_s}$ turns out to be $\gamma$-traceless and divergence free. Note also that, in view of \rf{26112013-man01-41}, conformal dimension of the spin-$(s + \half)$ field $\psi_{\cur,\d}^{a_1\ldots a_s}$ is equal to $s+d-\frac{3}{2}$. This implies that our gauge invarint approach to conformal current is equivalent to the standard CFT.

\newsection{ AdS/CFT correspondence for non-normalizable
modes of massless AdS field and shadow field}\label{secAdS/CFTsh}

We now discuss the AdS/CFT correspondence for bulk spin-$(s+\half)$ massless AdS field and boundary spin-$(s+\half)$ shadow field. We begin with an analysis of the non-normalizable solution of Eq.\rf{22112013-man1-09}. Solution of Eq.\rf{22112013-man1-09} with the
Dirichlet problem corresponding to the spin-$(s+\half)$ shadow field takes the form
\beq
\label{23112013-man1-11} |\psi(x,z)\rangle  &  =  &  n_\nu \int d^dy\, G_\nu(x-y,z) |\psi_\sh(y)\rangle\,,
\\
\label{24112013-man1-05} && n_\nu \equiv   \frac{(-)^{N_z}\Gamma(\nu)}{2^{N_z + \pi_-}\Gamma(\nu_s)} \,,
\eeq
where the Green function is given in \rf{22112013-man1-015}, while
$\nu_s$ is defined in \rf{26112013-man01-05}.

Using asymptotic behavior of the Green function $G_\nu$ \rf{22112013-man1-016}, we
find the asymptotic behavior of our solution
\be \label{man01-15102011-48}
|\psi(x,z)\rangle  \,\,\, \stackrel{z\rightarrow 0 }{\longrightarrow}\,\,\,
z^{-\nu + \half} n_\nu |\psi_\sh(x)\rangle\,.
\ee
From this expression, we see that solution \rf{23112013-man1-11} has indeed
asymptotic behavior corresponding to the spin-$(s+\half)$ shadow field.%
\footnote{ Since solution \rf{23112013-man1-11} has nonintegrable asymptotic
behavior \rf{man01-15102011-48}, such solution is sometimes referred to as
the non-normalizable solution.}

Note that representation of AdS fields as 2-vectors in \rf{26112013-man01-01}  implies the corresponding representation of fields of spin-$(s+\half)$ shadow field,
\be \label{26112013-man01-36}
\psi_\sh^{a_1\ldots a_{s'}\alpha} = \left(
\begin{array}{c}
\psi_{\sh,\u}^{a_1\ldots a_{s'}\alpha}
\\[7pt]
\psi_{\sh,\d}^{a_1\ldots a_{s'}\alpha}
\end{array}
\right)\,, \qquad s'=0,1,\ldots, s\,.
\ee
Using representation of dilatation symmetry on space of AdS field in \rf{conalggenlis03} and solution in \rf{23112013-man1-11}, we find realization of the operator of conformal dimension on space of $|\psi_\sh\rangle$,
\be \label{26112013-man01-37}
\Delta_\sh =  2- s + N_z - \half \sigma_3\,.
\ee
From \rf{26112013-man01-37}, we see that conformal dimensions of fields in \rf{26112013-man01-36} are given by
\be
\Delta_\sh(\psi_{\sh,\u}^{a_1\ldots a_{s'}\alpha}) = \frac{3}{2} - s'\,, \qquad
\Delta_\sh(\psi_{\sh,\d}^{a_1\ldots a_{s'}\alpha}) = \frac{5}{2}-  s'\,.
\ee
Note that the choice of normalization factor $n_\nu$ in  \rf{23112013-man1-11} is
a matter of convenience. Our normalization condition implies the following
normalization of asymptotic behavior of the solution for leading rank-$s$ tensor-spinor field in \rf{man01-15102011-48},
\be
\psi_\u^{a_1\ldots a_s}(x,z) \ \ \stackrel{z\rightarrow 0}{\longrightarrow} \ \  z^{-\nu_s + \half}
\psi_{\sh,\u}^{a_1\ldots a_s }(x)\,,
\ee
where $\nu_s$ is given in \rf{26112013-man01-05}.

Now, we are going to prove the following statements:

\noindent {\bf i}) For solution \rf{23112013-man1-11}, the first-order equations of motion \rf{12072013-09} and modified de Donder gauge condition \rf{22112013-man1-08} lead to differential constraints for the shadow field.

\noindent {\bf ii}) On-shell leftover gauge transformation
\rf{22112013-man1-06} of solution \rf{23112013-man1-11} leads to gauge
transformation of the spin-$(s+\half)$ shadow field.

\noindent {\bf iii}) action evaluated on
solution \rf{23112013-man1-11}  coincides, up to normalization
factor, with boundary two-point gauge invariant vertex for the
shadow field.

Below we demonstrate how these statements can be proved by using the
following relations for the Green function $G_\nu \equiv G_\nu(x-y,z)$:
\beq
\label{man01-01112010-24} && \TT_{-\nu+\half}G_{\nu-1} = - 2(\nu-1) G_\nu\,,
\\
\label{man01-01112010-24a1} && \TT_{\nu+\half}G_{\nu+1} = \frac{1}{2\nu} \Box G_\nu\,.
\eeq

{\bf Matching of bulk modified de Donder gauge and boundary constraint}. We
now demonstrate how differential constraints for the shadow field are obtained from first-order equations of motion \rf{12072013-09} and modified de Donder gauge condition \rf{22112013-man1-08}. To this end we note the relations
\beq
\label{24112013-man1-14} && e_1 (n_\nu G_\nu)  =  (n_\nu G_\nu) (\overset{\leftarrow}\Box{}_y \alpha^z\ewt_1) \,,
\\
\label{24112013-man1-15} && \bar{e}_1 (n_\nu G_\nu)  = (n_\nu G_\nu ) (-\ewt_1\bar\alpha^z)\,,
\\
\label{24112013-man1-16} && e_1^\Gammasm  (n_\nu G_\nu)  = (n_\nu G_\nu) e_{1,1}^\Gammasm(\sigma_+ + \sigma_- \overset{\leftarrow}\Box{}_y) \,,
\eeq
where Laplace operator $\overset{\leftarrow}\Box{}_y\equiv \frac{\overset{\leftarrow}\partial}{\partial y^a} \frac{\overset{\leftarrow}\partial}{\partial y^a}$ appearing in \rf{24112013-man1-14} and \rf{24112013-man1-16} is acting on the Green function $G_\nu = G_\nu(x-y,z)$.
Acting with operators $\Ewh$ \rf{12072013-01} and $\Cb_\mod^\on-sh$ \rf{23112013-man1-14} on solution $|\psi\rangle$ \rf{23112013-man1-11} and using \rf{24112013-man1-14}-\rf{24112013-man1-16}, we obtain the relations
\beq \label{man01-15102011-50x}
&& \Ewh |\psi\rangle  =  n_\nu \int d^dy\, G_\nu(x-y,z) \Ewh_\sh
|\psi_\sh(y)\rangle\,,
\\
\label{man01-15102011-50}
&& \Cb_\mod^\on-sh  |\psi\rangle  =  n_\nu \int d^dy\, G_\nu(x-y,z)\Cb_\sh
|\psi_\sh(y)\rangle\,,
\eeq
where operators $\Ewh_\sh$, $\Cb_\sh$ take the form
\beq
\label{24112013-man1-07} && \hspace{-1cm}  \Ewh_\sh \equiv \parline - G_\sh \gaalb + \Bigl( 1 + \gaal \frac{2}{2N_\alpha+d-2}\gaalb\Bigr)e_{1,\sh}^\Gammasm -\gamma\alpha \frac{2}{2N_\alpha+d-2}\eb_{1,\sh}\,,
\\
\label{24112013-man1-08} && \hspace{-1cm} \Cb_\sh \equiv \Cb_\st - \frac{1}{2N_\alpha+d-2}(\gaalb + \half \gaal \bar\alpha^2) e_{1,\sh}^\Gammasm + \half \bar\alpha^2 e_{1,\sh} - \frac{2N_\alpha+d-4}{2N_\alpha+d-2}\Pi_\bos^\smponetwo \eb_{1,\sh},\qquad
\\
\label{23112013-man1-15-n2} && G_\sh \equiv \alpha\partial - e_{1,\sh} + \gaal
\frac{1}{2N_\alpha+d-2}e_{1,\sh}^\Gammasm - \alpha^2 \frac{1}{2N_\alpha+d}\eb_{1,\sh}\,,
\\
&& \Cb_\st \equiv \albpar - \half \alpar\bar\alpha^2\,,
\\
\label{24112013-man1-09} && e_{1,\sh}^\Gammasm =  e_{1,1}^\Gammasm \bigl( \sigma_-  \Box + \sigma_+ \bigr)\,,
\\
\label{24112013-man1-10} && e_{1,\sh} = \Box \alpha^z \ewt_1\,,  \qquad \eb_{1,\sh} =  -   \ewt_1 \bar\alpha^z\,,
\\
\label{24112013-man1-11} && e_{1,1}^\Gammasm = \frac{2s+d-2}{2s+d-2-2N_z}\,, \qquad \ewt_1 = \Bigl(\frac{2s+d-3-N_z}{2s+d-4-2N_z}\Bigr)^{1/2}\,.
\eeq
From \rf{man01-15102011-50}, we see that first-order equations of motion \rf{12072013-09} and our modified de Donder gauge condition \rf{22112013-man1-08} lead indeed to the differential constraint for the shadow field given by
\beq
\label{26112013-man01-19} && \Ewh_\sh |\psi_\sh\rangle = 0\,,
\\
\label{26112013-man01-20} && \Cb_\sh |\psi_\sh\rangle = 0\,.
\eeq
As a side of remark we note that constraint \rf{26112013-man01-19} can be represented as
\beq
\label{27112013-man01-05} && E_\sh |\psi_\sh\rangle = 0\,,
\\
\label{27112013-man01-06} && E_\sh \equiv E_{\sh, \smone} + E_{\sh,\smzero}
\\
\label{27112013-man01-07} &&  E_{\sh,\smone} \equiv  E_\smone\,,
\\
\label{27112013-man01-08} && E_{\sh,\smzero} \equiv (1 - \gamma\alpha\gamma\bar\alpha -
\frac{1}{4}\alpha^2\bar\alpha^2) e_{1,\sh}^\Gammasm + (\gamma\alpha -
\half\alpha^2\gamma\bar\alpha) \eb_{1,\sh}
+ (\gamma\bar\alpha -\half \gamma\alpha\bar\alpha^2)  e_{1,\sh}\,,\qquad
\eeq
where the Fang-Fronsdal operator $E_\smone$ appearing in \rf{27112013-man01-07} is given in \rf{22112013-man1-05-c1}.

{\bf Matching of bulk and boundary gauge symmetries}. We now show how gauge transformation of the shadow field is obtained from the on-shell leftover
gauge transformation of the massless AdS field. To this end we note
that the corresponding on-shell leftover gauge transformation of massless AdS field is
obtained from \rf{22112013-man1-06} by plugging non-normalizable solution of equation for
gauge transformation parameter \rf{23112013-man1-16} into \rf{22112013-man1-06}. The
non-normalizable solution of equation \rf{23112013-man1-16} is given by
\be \label{man01-15102011-56}
|\xi(x,z)\rangle  =  n_\nu \int d^dy\,
G_\nu(x-y,z) |\xi_\sh(y)\rangle\,,
\ee
where $n_\nu$ is given in \rf{24112013-man1-05}. We now note that, on the one
hand, plugging \rf{man01-15102011-56} into \rf{22112013-man1-06} and using
\rf{23112013-man1-11}, we can cast the on-shell leftover gauge
transformation of $|\psi\rangle$ into the form
\be \label{man01-15102011-57}
\delta |\psi\rangle  =  n_\nu\int d^dy\, G_\nu(x-y,z) G_\sh
|\xi_\sh(y)\rangle\,,
\ee
where $G_\sh$ is given in \rf{23112013-man1-15-n2} . On the other hand, making use of relation \rf{23112013-man1-11}, we get
\be \label{man01-15102011-58}
\delta \psik   =  n_\nu \int d^dy\, G_\nu (x-y,z) \delta
|\psi_\sh(y)\rangle\,.
\ee
Comparing \rf{man01-15102011-57} with \rf{man01-15102011-58}, we obtain gauge transformation of the shadow field,
\be \label{26112013-man01-40x}
\delta |\psi_\sh\rangle =  G_\sh|\xi_\sh\rangle\,.
\ee
We check that differential constraints \rf{26112013-man01-19}, \rf{26112013-man01-20} are invariant under gauge transformation \rf{26112013-man01-40x}. Thus we see that the on-shell
leftover gauge symmetries of solution of the Dirichlet problem for the spin-$(s+\half)$ massless AdS field are indeed lead to gauge symmetries of the spin-$(s+\half)$ shadow field.

{\bf Matching of effective action and boundary two-point vertex}. To find the
effective action we follow the standard procedure. Namely, we plug
non-normalizable solution of the bulk equation of motion with the Dirichlet
problem corresponding to the boundary shadow field
\rf{23112013-man1-11} into the bulk action for AdS massless field. In other words, we are going to compute an effective action defined by the relation%
\footnote{ In this paper, for the study of  AdS/CFT correspondence for massless fields, we use Lagrangian approach. Study of the  AdS/CFT correspondence by using equations of motion and higher-spin symmetries may be found e.g., in Refs.\cite{Giombi:2009wh}-\cite{Didenko:2013bj}.}
\be \label{24112013-man1-17}
\Gamma_\eff \equiv \int d^dxdz\,  \LL_{\rm on-shell}\,,
\ee
where $ \LL_{\rm on-shell}$ is the Lagrangian evaluated on non-normalizable solution of the bulk equation of motion with the Dirichlet problem corresponding to the boundary shadow field
\rf{23112013-man1-11}. Note however that the Lagrangian presented in \rf{22112013-man1-05} does not involve a proper boundary term. We find the following expression for Lagrangian involving the proper boundary term:%
\footnote{ We note that boundary terms proportional to $e_{1,1}^\Gammasm$  in \rf{24112013-man1-18} can be fixed by using the same methods as for spin-$\half$ field in Refs.\cite{Arutyunov:1998ve,Henneaux:1998ch}. After this, the remaining boundary terms which are proportional to $e_{1,1}$ and $\eb_{1,1}$ can simply be fixed by requiring the $\Gamma_\eff$ to be invariant under gauge transformation in \rf{26112013-man01-40x}. For the case of bosonic fields, interesting discussion of gauge symmetries of boundary terms may be found in Ref.\cite{Joung:2011xb}.}
\beq
\label{24112013-man1-18} && \hspace{-1cm} {\rm i} \LL =   \psibr \overset{\leftrightarrow}E \psik\,,
\\
&&  \overset{\leftrightarrow}E \equiv  \overset{\leftrightarrow}E_\smone +  \overset{\leftrightarrow}E_\smzero\,,
\\
&&  \overset{\leftrightarrow}E_\smone \equiv
\overset{\leftrightarrow}\parline  -  \alpha\overset{\leftrightarrow}\partial\gamma\bar\alpha -
\gamma\alpha\bar\alpha\overset{\leftrightarrow}\partial + \gamma\alpha
\overset{\leftrightarrow}\parline\gamma\bar\alpha + \frac{1}{2}\gamma\alpha\alpha\overset{\leftrightarrow}\partial\bar\alpha^2
+ \frac{1}{2}\alpha^2\gamma\bar\alpha\bar\alpha\overset{\leftrightarrow}\partial -
\frac{1}{4}\alpha^2\overset{\leftrightarrow}\parline\bar\alpha^2\,, \ \ \ \
\\
&& \overset{\leftrightarrow}E_\smzero = (1 - \gamma\alpha\gamma\bar\alpha -
\frac{1}{4}\alpha^2\bar\alpha^2) \overset{\leftrightarrow}e{}_1^\Gammasm + (\gamma\alpha -
\half\alpha^2\gamma\bar\alpha)\overset{\leftarrow}\eb{}_1
+ (\gamma\bar\alpha -\half \gamma\alpha\bar\alpha^2) \overset{\rightarrow}e{}_1\,,
\eeq

\beq
&& \overset{\leftrightarrow}e{}_1^\Gammasm =  e_{1,1}^\Gammasm \bigl( \sigma_- \overset{\rightarrow}\TT_{\nu-\half} + \overset{\leftarrow}\TT_{\nu-\half} \sigma_+\bigr)\,,
\\
&& \overset{\rightarrow}e{}_1 = e_{1,1} \overset{\rightarrow}\TT_{\nu-\half}\,, \qquad \overset{\leftarrow}\eb_1 =  -\overset{\leftarrow}\TT_{\nu - \half} \eb_{1,1}\,,
\\
&& e_{1,1} = -\alpha^z \ewt_1 \,, \qquad \eb_{1,1} =  - \ewt_1 \bar\alpha^z\,,\qquad \ewt_1 = \Bigl(\frac{2s+d-3-N_z}{2s+d-4-2N_z}\Bigr)^{1/2}\,,
\\
&& e_{1,1}^\Gammasm = \frac{2s+d-2}{2s+d-2-2N_z}\,,
\\
\hspace{1cm} && \overset{\leftrightarrow}\partial{}^a \equiv \half (\overset{\rightarrow}\partial{}^a -  \overset{\leftarrow}\partial{}^a)\,, \qquad \overset{\rightarrow}\TT{}_\nu \equiv \overset{\rightarrow}\partial{}_z +
\frac{\nu}{z}\,,  \qquad \overset{\leftarrow}\TT{}_\nu \equiv \overset{\leftarrow}\partial{}_z +
\frac{\nu}{z}\,,
\\
\label{26112013-man01-15} && \nu \equiv s + \frac{d-3}{2} - N_z + \half \sigma_3\,.
\eeq
It is easy to see that Lagrangian \rf{24112013-man1-18} differs from the one in \rf{22112013-man1-05} by total derivatives $\partial^a$, $\partial_z$.

Lagrangian \rf{24112013-man1-18} evaluated on the solution of first-order equations of motion \rf{12072013-04} takes the form
\beq
\label{26112013-man01-06} && \irm \LL_{\rm on-shell} = \half \partial_z \Bigl( \langle \psi(x,z)| E_\eff |\psi(x,z) \rangle\Bigr) \,,
\\
\label{26112013-man01-07} && E_\eff \equiv (1 - \gamma\alpha\gamma\bar\alpha -
\frac{1}{4}\alpha^2\bar\alpha^2) e_{1,1}^\Gammasm \sigma_1 - (\gamma\alpha -
\half\alpha^2\gamma\bar\alpha)\eb_{1,1}
+ (\gamma\bar\alpha -\half \gamma\alpha\bar\alpha^2)  e_{1,1}\,.
\eeq
Taking into account \rf{24112013-man1-17} we obtain the following expression  for the effective action:
\be \label{26112013-man01-08}
-\irm \Gamma_\eff \equiv \half \int d^dx \langle \psi(x,z)| E_\eff |\psi(x,z) \rangle \Bigr|_{z\rightarrow 0}\,.
\ee
Now, plugging in \rf{26112013-man01-08} solution to the second-order gauge-fixed equations of motion \rf{23112013-man1-11}, we obtain the following three equivalent representations for the effective action:

\noindent {\bf 1st representation for the effective action}:
\beq
\label{26112013-man01-09} && \hspace{-1cm} - \irm \Gamma_\eff  =  \nu_s c_{\nu_s} \int d^d x_1d^d x_2 \langle \psi_\sh(x_1)| \Bigl( (\sigma_+ + (\gaal - \half\alpha^2\gaalb) \eb_{1,1}) \frac{f_\nu}{|x_{12}|^{2\nu+d}}
\nonumber\\
&& \hspace{3cm} + \, \frac{f_\nu}{|x_{12}|^{2\nu+d}} (\sigma_-  - (\gaalb - \half\gaal \bar\alpha^2) e_{1,1}) \Bigr)|\psi_\sh(x_2)\rangle\,,
\\
&& f_\nu \equiv \frac{\Gamma(\nu+1)\Gamma(\nu+\frac{d}{2})}{ 4^{\nu_s - \nu} \Gamma(\nu_s+1)\Gamma(\nu_s+\frac{d}{2})}\,,
\\
\label{26112013-man01-10} && \nu_s \equiv s + \frac{d-2}{2}\,,
\eeq
where $c_\nu$ and $\nu$ are given in \rf{22112013-man1-015-b5} and \rf{26112013-man01-15} respectively;

\noindent {\bf 2nd representation for the effective action}:
\beq
\label{26112013-man01-11} && \hspace{-1cm} - \irm \Gamma_\eff  = \frac{c_{\nu_s}}{4(\nu_s+\frac{d}{2}-1)} \int d^d x_1d^d x_2 \langle \psi_\sh(x_1)| \frac{g_\nu}{|x_{12}|^{2\nu+d-2}}E_{\sh,\smzero} |\psi_\sh(x_2)\rangle\,
\\
\label{26112013-man01-12} && g_\nu \equiv \frac{\Gamma(\nu)\Gamma(\nu+\frac{d}{2}-1)}{ 4^{\nu_s - \nu} \Gamma(\nu_s)\Gamma(\nu_s+\frac{d}{2}-1)}\,,
\\
\label{26112013-man01-14} && E_{\sh,\smzero} \equiv (1 - \gamma\alpha\gamma\bar\alpha -
\frac{1}{4}\alpha^2\bar\alpha^2) e_{1,\sh}^\Gammasm + (\gamma\alpha -
\half\alpha^2\gamma\bar\alpha) \eb_{1,\sh}
+ (\gamma\bar\alpha -\half \gamma\alpha\bar\alpha^2)  e_{1,\sh}\,.\qquad
\eeq

Effective action \rf{26112013-man01-09}, \rf{26112013-man01-11} is invariant under gauge transformation \rf{26112013-man01-40x} provided the shadow field $|\psi_\sh\rangle$  satisfies the differential constraints \rf{26112013-man01-19}, \rf{26112013-man01-20}. Note also that using constraint \rf{27112013-man01-05}, we can represent \rf{26112013-man01-11} in terms of the Fang-Fronsdal operator.

\noindent {\bf 3rd representation for the effective action}:
\be \label{28112013-man01-01}
\irm \Gamma_\eff  = \frac{c_{\nu_s}}{4(\nu_s+\frac{d}{2}-1)} \int d^d x_1d^d x_2 \langle \psi_\sh(x_1)| \frac{g_\nu}{|x_{12}|^{2\nu+d-2}} E_{\sh,\smone} |\psi_\sh(x_2)\rangle\,,
\ee
where $E_{\sh,\smone} = E_\smone$ and the Fang-Fronsdal operator $E_\smone$ is given in \rf{22112013-man1-05-c1}.

To summarize, using CFT adapted action and the modified de Donder gauge, we obtain the 2-point gauge invariant effective action for the spin-$(s+\half)$ shadow field.  Representations for the effective action given in \rf{26112013-man01-09}, \rf{26112013-man01-11}, \rf{28112013-man01-01} is our solution to the problem of 2-point effective action for the case of fermionic arbitrary spin fields.%
\footnote{ In this paper, we discuss 2-point effective action. Recent results on Lagrangian description of interacting AdS fields (see e.g., Refs.\cite{Metsaev:2006ui}-\cite{Henneaux:2012wg}) provides  interesting possibility for the studying interaction dependent contributions to the effective action.}
Our effective action is gauge invariant under gauge transformation \rf{26112013-man01-40x} and expressed in terms of gauge fields \rf{26112013-man01-36} which are subject to the differential constraints \rf{26112013-man01-19}, \rf{26112013-man01-20}. By fixing the gauge symmetries in various ways, we can obtain various new representations for the effective action. For instance, we can use Stueckelberg gauge frame or light-cone gauge frame. In Stueckelberg gauge frame, gauging away Stueckelberg fields and solving the differential constraints \rf{26112013-man01-19}, \rf{26112013-man01-20}, we can express all fields in \rf{26112013-man01-36} in terms of the $\gamma$-traceless rank-$s$ tensor-spinor field of the $so(d-1,1)$ algebra. This $\gamma$-traceless field turns out to be free of differential constraints.%
\footnote{ Note that the fact that we can express all fields in \rf{26112013-man01-36} in terms of the one rank-$s$ $\gamma$-traceless tensor-spinor field of the $so(d-1,1)$ algebra implies that our gauge invariant approach is equivalent to the standard approach to CFT.}
Plugging such solution for fields \rf{26112013-man01-36} into \rf{26112013-man01-09}, \rf{26112013-man01-11}, \rf{28112013-man01-01} provides the representation for the effective action in terms of the $\gamma$-traceless rank-$s$ tensor-spinor field. Such representation gives 2-point function of shadow field in the standard CFT.
In light-cone gauge frame, using light-cone gauge and the differential constraints \rf{26112013-man01-19}, \rf{26112013-man01-20}, we can express all our fields \rf{26112013-man01-36} in terms of one rank-$s$ light-cone tensor-spinor fields. Plugging such solution for fields \rf{26112013-man01-36} into \rf{26112013-man01-09}, \rf{26112013-man01-11}, \rf{28112013-man01-01} provides light-cone gauge representation for the effective action. In other words, one of advantages of our
approach is that our approach gives the possibility for the studying the effective action by using various  gauge conditions which might be preferable in various applications.

\section{Conformal fermionic fields}\label{man02-sec-07}

The kernel of effective action given in \rf{26112013-man01-09}, \rf{26112013-man01-11}, \rf{28112013-man01-01} is not well-defined when $d$
is even integer and $\nu$ takes integer values (see e.g.
\cite{Aref'eva:1998nn}). However this kernel can be regularized and after
that it turns out that the leading logarithmic divergence of the effective action
$\Gamma_\eff$ leads to Lagrangian of conformal fermionic fields. To explain what has
just been said we note that the kernel of $\Gamma_\eff$ can be regularized by
using dimensional regularization. This is to say that using the dimensional
regularization and denoting the integer part of $d$ by $[d]$, we introduce
the regularization parameter $\epsilon$ as
\be  \label{man02-20072009-29}
d - [d] = - 2\epsilon\,,\qquad [d]-\hbox{even
integer} \,.
\ee
With this notation we note that $\nu$ \rf{26112013-man01-15} can be presented as%
\footnote{ Operator $[\nu]$ in \rf{26112013-man01-16} takes the form $[\nu] = s - N_z - \frac{3}{2} + \half \sigma_3$. Obviously this operator has integer eigenvalues.}
\be \label{26112013-man01-16}
\nu = [\nu] + \frac{d}{2} \,, \qquad  [\nu] - \hbox{integer} \,.
\ee
Now we use the following well know fact. With $d$ and $\nu$ given in \rf{man02-20072009-29} and \rf{26112013-man01-16} respectively the regularized
kernel in \rf{26112013-man01-11} has the following behavior:
\beq \label{man02-20072009-30}
&& \frac{1}{|x|^{2\nu+d-2}}\,\,\, \stackrel{\epsilon \sim
0}{\mbox{\Large$\sim$}}\,\,\, \frac{1}{\epsilon} \varrho_{\nu-1} \Box^{\nu-1}
\delta^{(d)}(x)\,,
\\[5pt]
&&  \label{man02-20072009-31} \varrho_\nu = \frac{\pi^{d/2}}{4^\nu \Gamma(\nu
+ 1)\Gamma(\nu + \frac{d}{2})}\,.
\eeq
Using \rf{man02-20072009-30} in
\rf{26112013-man01-11}, we obtain
\be  \label{man02-20072009-32}
\Gamma_\eff \,\,\, \stackrel{ \epsilon \sim
0}{\mbox{\Large$\sim$}}\,\,\, \frac{1}{\epsilon} c_{\nu_s} {\nu_s}\varrho_{\nu_s} \int
d^dx\,\, \LL\,, \ee
where $\nu_s$ is defined in \rf{26112013-man01-10} and $\LL$ is a
higher-derivative Lagrangian for conformal spin-$(s+\half)$ fermionic field. The Lagrangian takes the form
\beq
\label{26112013-man01-17} && \hspace{-1cm}  \irm \LL =
\langle\psi_\cf| \Box^{\nu-1}E_{\cf,\smzero} |\psi_\cf\rangle\,,
\\
\label{26112013-man01-18} && E_{\cf,\smzero} \equiv (1 - \gamma\alpha\gamma\bar\alpha -
\frac{1}{4}\alpha^2\bar\alpha^2) e_{1,\cf}^\Gammasm + (\gamma\alpha -
\half\alpha^2\gamma\bar\alpha) \eb_{1,\cf}
+ (\gamma\bar\alpha -\half \gamma\alpha\bar\alpha^2)  e_{1,\cf}\,,
\\
&& e_{1,\cf}^\Gammasm =  e_{1,1}^\Gammasm \bigl( \sigma_-  \Box + \sigma_+ \bigr)\,,
\\
&& e_{1,\cf} = \Box \alpha^z \ewt_1\,,  \qquad \eb_{1,\cf} =  -   \ewt_1 \bar\alpha^z\,,
\\
&& e_{1,1}^\Gammasm = \frac{2s+d-2}{2s+d-2-2N_z}\,, \qquad \ewt_1 = \Bigl(\frac{2s+d-3-N_z}{2s+d-4-2N_z}\Bigr)^{1/2}\,,
\eeq
where we have made the identification for the ket-vector of the conformal field $|\psi_\cf\rangle$,
\be \label{man02-21072009-09}
|\psi_\cf\rangle   =  |\psi_\sh\rangle\,,
\ee
i.e., the ket-vector $|\psi_\cf\rangle$ is represented in terms of tensor-spinor components as
\beq
\label{26112013-man01-33} && |\psi_\cf\rangle  \equiv \sum_{s'=0}^s \frac{\alpha_z^{s-s'}}{\sqrt{(s-s')!}}
|\psi^{s'}\rangle\,,
\\
&&  |\psi^{s'}\rangle \equiv \frac{1}{s'!} \alpha^{a_1}\ldots
\alpha^{a_{s'}} \psi_\cf^{a_1\ldots a_{s'}\alpha}|0\rangle\,.
\eeq
Using this identification, we note that the differential constraints for the
spin-$(s+\half)$ shadow field given in \rf{26112013-man01-19}, \rf{26112013-man01-20} imply the same differential constraints for the conformal field $|\psi_\cf\rangle$,
\beq
\label{26112013-man01-21} && \Ewh_\cf |\psi_\cf\rangle = 0\,,
\\
\label{26112013-man01-22} && \Cb_\cf |\psi_\cf\rangle = 0\,,
\eeq
where
\beq
&& \hspace{-1cm}  \Ewh_\cf \equiv \parline - G_\cf \gaalb + \Bigl( 1 + \gaal \frac{2}{2N_\alpha+d-2}\gaalb\Bigr)e_{1,\cf}^\Gammasm -\gamma\alpha \frac{2}{2N_\alpha+d-2}\eb_{1,\cf}\,,
\\
&& \hspace{-1cm} \Cb_\cf \equiv \Cb_\st - \frac{1}{2N_\alpha+d-2}(\gaalb + \half \gaal \bar\alpha^2) e_{1,\cf}^\Gammasm + \half \bar\alpha^2 e_{1,\cf} - \frac{2N_\alpha+d-4}{2N_\alpha+d-2}\Pi_\bos^\smponetwo \eb_{1,\cf}\,,
\\
&& \Cb_\st \equiv \albpar - \half \alpar\bar\alpha^2\,.
\eeq

Constraints \rf{26112013-man01-21}, \rf{26112013-man01-22} are invariant under the gauge transformation
\beq
\label{26112013-man01-23} && \delta |\psi_\cf\rangle = G_\cf |\xi_\cf\rangle\,,
\\
&& \hspace{1cm} G_\cf \equiv \alpha\partial - e_{1,\cf} + \gaal
\frac{1}{2N_\alpha+d-2}e_{1,\cf}^\Gammasm - \alpha^2
\frac{1}{2N_\alpha+d}\eb_{1,\cf} \,,
\eeq
where ket-vector $|\xi_\cf\rangle$ takes the same form as the ket-vector
$|\xi\rangle$ given in \rf{gaugpar1-a1}.%
\footnote{ Note however that ket-vector $|\xi\rangle$ given in \rf{gaugpar1-a1}
depends on $x^a$ and $z$, while the ket-vector $|\xi_\cf\rangle$ appearing in \rf{26112013-man01-23} depends only on $x^a$.}
Also it is easy  to check that Lagrangian \rf{26112013-man01-17} is invariant under gauge transformation \rf{26112013-man01-23} provided the ket-vector
$|\psi_\cf\rangle$ satisfies differential constraints \rf{26112013-man01-21}, \rf{26112013-man01-22}.

The following remarks are in order:

\noindent {\bf i)} Constraint \rf{26112013-man01-21} can be represented as
\beq
\label{27112013-man01-09} && E_\cf |\psi_\cf\rangle = 0\,,
\\
\label{27112013-man01-10} && E_\cf \equiv E_{\cf,\smone} + E_{\cf,\smzero}
\\
\label{27112013-man01-11} &&  E_{\cf,\smone} \equiv  E_\smone\,,
\eeq
where the Fang-Fronsdal operator $E_\smone$ appearing in \rf{27112013-man01-11} is given in \rf{22112013-man1-05-c1}, while the operator $E_{\cf,\smzero}$ is given in \rf{26112013-man01-18}. Using \rf{27112013-man01-09},  we see that Lagrangian \rf{26112013-man01-17}  can be represented in terms of the Fang-Fronsdal oprator,
\be
- \irm \LL = \langle\psi_\cf| \Box^{\nu-1}E_{\cf,\smone} |\psi_\cf\rangle\,.
\ee

\noindent {\bf ii)} Using differential constraints \rf{26112013-man01-21}, \rf{26112013-man01-22} and gauging away Stueckelberg fields we can obtain representation for all tensor-spinor fields appearing in \rf{26112013-man01-33} in terms of one rank-$s$ $\gamma$-traceless tensor-spinor field of $so(d-1,1)$ algebra which is not subject to any differential constraints. Plugging such representation for the tensor-spinor fields in Lagrangian \rf{26112013-man01-17}, we can express our Lagrangian in terms of the one rank-$s$  $\gamma$-traceless tensor-spinor field.

\noindent {\bf iii)} We note that UV divergence of the effective action leads to higher-derivative action of the conformal fields.%
\footnote{ Ordinary-derivative actions  for conformal fields are discussed in Refs.\cite{Metsaev:2007fq}-\cite{Metsaev:2012hr}.}

{\bf Acknowledgments}. This work was supported by the RFBR Grant
No.11-02-00685.

%%%%%%%%%%%%%%%%%%%%%%%%%%%%%%%%%%%%%%%%%%%%%%%%%%%%%%%%%%%%%%%%%%%%%%%%%
%%%%%%%%%%%%%%%%%%%%%%%%%%%%%%%%%%%%%%%%%%%%%%%%%%%%%%%%%%%%%%%%%%%%%%%%%
\setcounter{section}{0} \setcounter{subsection}{0}
\appendix{ Notation }
%%%%%%%%%%%%%%%%%%%%%%%%%%%%%%%%%%%%%%%%%%%%%%%%%%%%%%%%%%%%%%%%%%%%%%%%%
%%%%%%%%%%%%%%%%%%%%%%%%%%%%%%%%%%%%%%%%%%%%%%%%%%%%%%%%%%%%%%%%%%%%%%%%%

Vector indices of the $so(d-1,1)$ algebra  take the values $a,b,c=0,1,\ldots
,d-1$, while vector indices of the $so(d,1)$ algebra take the values
$A,B,C=0,1,\ldots ,d-1,d$. We use mostly positive flat metric tensors
$\eta^{ab}$, $\eta^{AB}$. To simplify our expressions we drop $\eta_{ab}$,
$\eta_{AB}$ in the respective scalar products, i.e., we use $X^a Y^a \equiv
\eta_{ab}X^a Y^b$, $X^A Y^A \equiv \eta_{AB}X^A Y^B$. Using the
identification $X^d \equiv X^z$ gives the following decomposition of the
$so(d,1)$ algebra vector: $X^A=X^a,X^z$. This implies $X^AY^A = X^aY^a +
X^zY^z$.

We use the creation operators $\alpha^a$, $\alpha^z$, and the respective
annihilation operators $\bar{\alpha}^a$, $\bar{\alpha}^z$,
\be
[\bar{\alpha}^a,\alpha^b]=\eta^{ab}\,, \qquad [\bar\alpha^z,\alpha^z]=1\,,
\qquad
\bar\alpha^a |0\rangle = 0\,,\qquad  \bar\alpha^z |0\rangle = 0\,.\ee
These operators are referred to as oscillators in this paper. The oscillators
$\alpha^a$, $\bar\alpha^a$ and $\alpha^z$, $\bar\alpha^z$, transform in the
respective vector and scalar representations of the $so(d-1,1)$ algebra and
satisfy the hermitian conjugation rules, $\alpha^{a\dagger} = \bar\alpha^a$,
$\alpha^{z\dagger} = \bar\alpha^z$. Oscillators  $\alpha^a$, $\alpha^z$ and
$\bar\alpha^a$, $\bar\alpha^z$ are collected into the respective $so(d,1)$
algebra oscillators $\alpha^A =\alpha^a,\alpha^z$ and $\bar\alpha^A
=\bar\alpha^a,\bar\alpha^z$.

$x^A = x^a,z $ denote coordinates in $d+1$-dimensional $AdS_{d+1}$ space,
\be \label{speech01}
ds^2 = \frac{1}{z^2}(dx^a dx^a + dz dz)\,,
\ee
while $\partial_A=\partial_a,\partial_z$ denote the respective derivatives,
$\partial_a \equiv \partial / \partial x^a$, $\partial_z \equiv
\partial / \partial z$.

We use $2^{[(d+1)/2]}\times 2^{[(d+1)/2]}$ Dirac gamma matrices $\gamma^A$ in
$d+1$-dimensions, $ \{ \gamma^A,\gamma^B\} = 2\eta^{AB}$, and adapt
the following hermitian conjugation rules for the derivatives,
oscillators, and $\gamma$-matrices:
\be
\partial^{A\dagger} = - \partial^A, \qquad
\gamma^{A\dagger} = \gamma^0 \gamma^A\gamma^0\,,\qquad
\alpha^{a\dagger} = \bar\alpha^a\,, \qquad \alpha^{z\dagger} = \bar\alpha^z
\,.
\ee
We use operators constructed out of the derivatives, oscillators, and
$\gamma$-matrices,
\beq \label{manold-31102011-02}
&& \Box \equiv \partial^a\partial^a\,, \qquad\quad
\parline\equiv \gamma^a\partial^a\,,\qquad\quad
\alpha\partial \equiv \alpha^a\partial^a\,,\qquad\quad \bar\alpha\partial
\equiv \bar\alpha^a\partial^a\,,\qquad
\\
\label{manold-31102011-03} && \gamma\alpha \equiv \gamma^a\alpha^a\,,\qquad\
\ \gamma\bar\alpha \equiv \gamma^a\bar\alpha^a\,,\qquad \ \ \
\alpha^2 \equiv \alpha^a\alpha^a\,, \qquad\quad   \bar\alpha^2 \equiv
\bar\alpha^a\bar\alpha^a\,,\qquad
\\
\label{manold-31102011-05} && N_\alpha \equiv \alpha^a \bar\alpha^a \,,
\qquad \
N_z \equiv \alpha^z \bar\alpha^z\,, \qquad\quad \ \ \
\\
\label{26112013-man01-03} && \Pi^\smponethree  = 1- \gaal \frac{1}{2N_\alpha+d}\gaalb
- \alpha^2 \frac{1}{2(2N_\alpha+d+2)} \bar\alpha^2 \,,
\\
\label{26112013-man01-04} && \Pi_\bos^\smponetwo  = 1 - \alpha^2 \frac{1}{2(2N_\alpha+d)} \bar\alpha^2 \,.
\eeq
The $2\times 2$ matrices and antisymmetric products of $\gamma$-matrices are
defined as
\beq \label{manold-03112011-01}
&& \sigma_+  = \left(
\begin{array}{ll}
0 & 1
\\
0 & 0
\end{array}\right),
\quad
\sigma_-  = \left(
\begin{array}{ll}
0 & 0
\\
1 & 0
\end{array}\right),
\quad
\pi_+  = \left(
\begin{array}{ll}
1 & 0
\\
0 & 0
\end{array}\right),
\quad
\pi_-  = \left(
\begin{array}{ll}
0 & 0
\\
0 & 1
\end{array}\right)\,,
\\
\label{manold-31102011-04} && \hspace{1cm} \gamma^{ab} = \half (\gamma^a \gamma^b -
\gamma^b \gamma^a)\,,\qquad \gamma^{abc} =
\frac{1}{3!}(\gamma^a\gamma^b\gamma^c \pm 5 \hbox{ terms})\,.
\eeq
Notation $\sigma_1$, $\sigma_2$, $\sigma_3$ stands for the standard Pauli matrices.

The covariant derivative $D^A$ is given by $D^A = \eta^{AB}D_B$,
\beq \label{vardef01}
&& D_A \equiv e_A^\mu D_\mu\,,  \qquad D_\mu \equiv
\partial_\mu
+\frac{1}{2}\omega_\mu^{AB}M^{AB}\,,
\\
&& M^{AB} \equiv \alpha^A \bar\alpha^B - \alpha^B \bar\alpha^A + \half \gamma^{AB}\,,\qquad \gamma^{AB}\equiv \frac{1}{2}(\gamma^A\gamma^B - \gamma^B\gamma^A)\,,
\eeq
$\partial_\mu = \partial/\partial x^\mu$, where $e_A^\mu$ is inverse vielbein
of $AdS_{d+1}$ space, $D_\mu$ is the Lorentz covariant derivative and the
base manifold index takes values $\mu = 0,1,\ldots, d$. The $\omega_\mu^{AB}$
is the Lorentz connection of $AdS_{d+1}$ space, while $M^{AB}$ is a spin
operator of the Lorentz algebra $so(d,1)$. Note that $AdS_{d+1}$ coordinates
$x^\mu$ carrying the base manifold indices are identified with coordinates
$x^A$ carrying the flat vectors indices of the $so(d,1)$ algebra, i.e., we
assume $x^\mu = \delta_A^\mu x^A$, where $\delta_A^\mu$ is Kronecker delta
symbol. $AdS_{d+1}$ space contravariant tensor-spinor field, $\Psi^{\mu_1\ldots
\mu_s}$, is related with field carrying the flat indices, $\Psi^{A_1\ldots
A_s}$, in a standard way $\Psi^{A_1\ldots A_s} \equiv e_{\mu_1}^{A_1}\ldots
e_{\mu_s}^{A_s} \Psi^{\mu_1\ldots \mu_s}$. Helpful commutators are given by
\be   [D^A,D^B]=\Omega^{ABC} D^C - M^{AB}\,,
\qquad [\bar\alphabf \Dbf, \alphabf \Dbf]= \Box_\AdS +
\frac{1}{2}M^{AB}M^{AB}\,,
\ee
where $\Omega^{ABC} = -\omega^{ABC}+\omega^{BAC}$ is a contorsion tensor and
we define $\omega^{ABC}\equiv e^{A\mu} \omega_\mu^{BC}$.

For the Poincar\'e parametrization of $AdS_{d+1}$ space, vielbein
$e^A=e^A_\mu dx^\mu$ and Lorentz connection, $de^A+\omega^{AB}\wedge e^B=0$,
are given by
\be\label{eomcho01} e_\mu^A=\frac{1}{z}\delta^A_\mu\,,\qquad
\omega^{AB}_\mu=\frac{1}{z}(\delta^A_z\delta^B_\mu
-\delta^B_z\delta^A_\mu)\,. \ee
With choice made in \rf{eomcho01}, the covariant derivative takes the form
$D^A= z \partial^A + M^{zA}$, $\partial^A=\eta^{AB}\partial_B$.

\end{document}